\documentclass[aps]{revtex4} 
\usepackage{psfig}
\usepackage{epsfig}
\usepackage{bm}

\newcommand{\be}{\begin{equation}}
\newcommand{\ee}{\end{equation}}
\newcommand{\bv}{{\bm v}}

\begin{document}

\title{Molecular dynamics simulations of vibrated granular gases}
 
\author{Alain Barrat\footnote{Electronic Address: Alain.Barrat@th.u-psud.fr} 
and Emmanuel Trizac\footnote{Electronic Address: 
Emmanuel.Trizac@th.u-psud.fr}
}

\affiliation{
Laboratoire de Physique Th{\'e}orique
(UMR 8627 du CNRS), B{\^a}timent 210, Universit{\'e} de
Paris-Sud, 91405 Orsay Cedex, France 
}
 
\date{\today}

\begin{abstract}

We present molecular dynamics simulations of mono- or bidisperse
inelastic granular gases driven by vibrating walls, in two dimensions
(without gravity).
Because of the energy injection at the boundaries, a situation often met
experimentally, density and temperature fields display heterogeneous
profiles in the direction perpendicular to the walls. 
A general equation of state for an arbitrary mixture of fluidized inelastic
hard spheres is derived and successfully tested against numerical data.
Single-particle velocity distribution functions with non-Gaussian features
are also obtained, and the influence of various parameters
(inelasticity coefficients, density\dots)
analyzed. The validity of a recently proposed Random Restitution
Coefficient model is assessed through the study of projected collisions
onto the direction perpendicular to that of energy injection.
For the binary mixture, the non-equipartition of translational 
kinetic energy is studied and
compared both to experimental data and to the case of homogeneous energy
injection (``stochastic thermostat''). The rescaled velocity 
distribution functions are found to be
very similar for both species.
\end{abstract} 

\maketitle 

\section{Introduction}

Due to the intrinsic dissipative character of inter-particle collisions,
an energy supply is requested to fluidize a granular gas. This is often 
achieved by a vibrating boundary, and the resulting vibro-fluidized
beds provide non trivial realizations of non equilibrium steady states. 
The understanding of such far from equilibrium systems requires a correct 
description of the energy exchange between the vibrating piston 
and the granular medium, as well as a macroscopic continuum theory to describe
the evolution of the relevant coarse-grained fields \cite{Brey1,Brey2}
(density, temperature etc...).
In particular, the derivation of an accurate equation of state is a key 
step in the hydrodynamic approach. 

A simple, fair and much studied theoretical framework to capture the 
inelastic nature of grain-grain collisions in a rapid granular flow 
is provided by the inelastic hard sphere model \cite{Goldhirsch,Kadanoff}. 
In this article, we present 
the results of extensive molecular dynamics (MD) simulations of inelastic
hard spheres driven by an energy injection
at the boundaries, for both a one component fluid (mono-disperse case) and 
a binary mixture (bidisperse situation). We analyze in detail the effects of
several parameters that may be difficult to tune experimentally, with 
a particular emphasis on the profiles of the hydrodynamic fields. 

This article is organized as follows: in section
\ref{sec:eos}, we present the model
and derive an equation of state for an arbitrary mixture of
inelastic hard spheres, going beyond the ideal gas contribution 
in view of performing 
accurate hydrodynamic tests. The equation of state obtained is a natural
generalization of its standard counterpart for elastic hard spheres.
The two following sections (\ref{sec:one} 
and \ref{sec:binary}) are
then devoted to molecular dynamics simulations for one-component systems
and for binary mixtures. In both cases, we restrict ourselves to
two-dimensional simulations, both for simplicity and for
comparisons with 2D experimental data 
\cite{Rouyer,menon,Rouyercont,Feitosa}. 
As in the experiments, the energy loss due to inelastic collisions 
is compensated
for by an energy injection by vibrating or thermal walls, which leads
to heterogeneous density and temperature profiles. The various
profiles and velocity distribution functions are studied and
compared with experiments whenever possible. Moreover, 
projecting the dynamics onto the direction perpendicular to that of energy
injection
allows to assess the validity of the
random restitution coefficient model proposed in 
\cite{RRC1,RRC2}. The influence of various parameters on the
non-equipartition of energy in a binary mixture is studied
in section \ref{sec:binary}, and comparison with experimental
data and with the case of homogeneous energy injection is performed. 
In this latter case, the velocity distribution functions are analyzed and shown
to be very similar for the two species.
Conclusions are finally presented in 
section \ref{sec:concl}.

\section{The model -- Computation of an equation of state}
\label{sec:eos}

We consider a mixture of ${\cal N}_s$ species of hard spheres in 
dimension $d$, 
with diameters $\sigma_i$ and masses $m_i$, where $1 \leq i \leq {\cal N}_s$. 
A binary collision between grains of species $i$ and $j$ is momentum
conserving and dissipates kinetic energy.
In the simplest version of the model, the collision $i$-$j$ is characterized
by one inelasticity parameter: 
the coefficient of normal restitution $\alpha_{ij}$. 
Accordingly, the pre-collisional velocities
(${\bm v}_i, \bv_j$) are transformed into the post-collisional couple 
(${\bm v}'_i, {\bm v}'_j$) such that
\begin{eqnarray}
\bm{v}_i' \, =\,  \bm{v}_i - \frac{m_j}{m_i+m_j} (1+\alpha_{ij})
(\widehat{\bm{\sigma}}\cdot \bm{v}_{ij})\widehat{\bm{\sigma}} 
\label{eq:coll1}\\
\bm{v}_j'\, = \, \bm{v}_j + \frac{m_i}{m_i+m_j} (1+\alpha_{ij})
(\widehat{\bm{\sigma}}\cdot \bm{v}_{ij})\widehat{\bm{\sigma}}
\label{eq:coll2}
\end{eqnarray}  
where $\bv_{ij}=\bv_i-\bv_j$ and $\widehat{\bm\sigma}$ is the center to center
unit vector from particle $i$ to $j$. Note that
$\alpha_{ij}=\alpha_{ji}$ to ensure the conservation of total
linear momentum $m_i \bv_i+m_j\bv_j$. 

We also considered an extension of the previous model allowing for
rotations, introducing a coefficient of tangential restitution $\alpha^t_{ij}$
\cite{Luding},
see appendix \ref{app:rot}.
The collision law (\ref{eq:coll1})-(\ref{eq:coll2}) is then recovered for
$\alpha_{ij}^t = -1$.  

Irrespective of the value of the tangential restitution coefficient $\alpha^t$,
the linear-momentum change for particle $i$ in a collision $i$-$j$ reads
\be
\delta {\bm p}_i \,=\, -\delta {\bm p}_j \,=\, 
-\frac{m_i\,m_j}{m_i+m_j} (1+\alpha_{ij}) 
(\widehat{\bm{\sigma}}\cdot \bm{v}_{ij})\widehat{\bm{\sigma}} .
\ee
In appendix \ref{app:eos}, we use this relation to compute an 
equation of state for the homogeneous isotropic mixture, invoking 
the virial theorem (the pressure is defined kinetically from the
momentum transfer and does not follow from a statistical mechanics 
derivation). 
The total density is denoted
$\rho$ and the partial densities $\rho_i = x_i \rho$ (the number fractions
$x_i$ are such that $\sum_i x_i=1$). 
The temperature of species $i$ is $T_i$, defined from the mean kinetic
energy of subpopulation $i$: $ d T_i = \langle m_i v_i^2\rangle$.
Only for an elastic system is the energy equipartition $T_i = T, \forall i$ 
recovered 
\cite{losert,garzo,Wildman,menon,Duparcmeur,Clelland,Dahl,MontaneroShear,MontaneroHCS,Equipart,puglisi2,Pagnani}.
It is found in appendix \ref{app:eos} that the pressure in dimension $d$ reads
\be
P \,=\, \sum_{i} \rho_i T_i \,+\, \rho \eta\, 2^{d-1}\,\sum_{i,j} 
x_ix_j\, \frac{m_j}{m_i+m_j} \, (1+\alpha_{ij}) \,\, T_i \, \,
\frac{\sigma_{ij}^d}{\langle \sigma^d\rangle} \, \, \chi_{ij},
\label{eq:eos}
\ee
independently of $\alpha^t_{ij}$,
where $\sigma_{ij} = (\sigma_i+\sigma_j)/2$, $\langle\sigma^d\rangle = \sum_{i}
x_i \sigma_i^d$,
$\eta$ is the packing fraction (e.g. $\eta = \pi \rho\langle\sigma^3\rangle/6$
in three dimensions), and the $\chi_{ij}$ are the pair correlation functions
at contact. The latter --unknown-- quantities may be approximated by their
elastic counterparts (see \cite{Santos} for a general procedure to infer 
reliable pair correlation functions in a multi-component
$d$-dimensional hard-sphere fluid from the equation of state of the 
mono-disperse
system). In the following analysis, it will turn sufficient to 
include only the low density behaviour $\chi_{ij}=1$ to improve
upon the ideal equation of state 
$ P=P^{\hbox{\scriptsize ideal}}= \sum_{i} \rho_i T_i$,
that holds for $\rho \to 0$ only. We emphasize that no approximation
has been made on the single-particle velocity distribution in the derivation 
of Eq. (\ref{eq:eos}) (the key assumption is that the two-body
distribution function factorizes at contact in a product of the single-particle
distribution \cite{Pre2}). 
 
It is instructive to check the validity of our equation of state by considering
the elastic limit where $\alpha_{ij} =1$ and $T_i=T$. A straightforward
calculation (under the reasonable and often made assumption 
that $\chi_{ij}=\chi_{ji}$)
shows that the mass ratio simplifies and expression (\ref{eq:eos})
may be cast in the form
\be
\frac{P}{\rho T} \,=\,1\,+\, \eta\, 2^{d-1}\,\sum_{i,j} 
x_i x_j\, \,\frac{\sigma_{ij}^d}{\langle \sigma^d\rangle} \, \, \chi_{ij},
\ee
which is the correct result (see e.g. \cite{Zhang}). In particular, for 
the single species (mono-disperse) problem, we recover the virial
relation $P/(\rho T) = 1 + 2^{d-1} \eta \, \chi$.

We finally generalize Eq. (\ref{eq:eos}) to the situation of a continuous
size distribution, with a probability density distribution $W(\sigma)$
(normalized to 1 so that $\langle \sigma^n\rangle = \int \sigma^n W$); 
the temperature is in general a continuous function $T(\sigma)$ of size and 
\be
\frac{P}{\rho} \,=\, \int W(\sigma) T(\sigma) \, d\sigma \,+\, 
\frac{\eta}{2}\,\,\int d\sigma\,d\sigma'\, W(\sigma)W(\sigma')
\, \frac{m_{\sigma'}}{m_\sigma+m_{\sigma'}} \, (1+\alpha_{\sigma{\sigma'}}) 
\,\, T(\sigma) \, \,
\frac{(\sigma+\sigma')^d}{\langle \sigma^d\rangle} \, \, 
\chi_{\sigma{\sigma'}}.
\ee
In the following sections, the above equation of state will be 
used to test hydrodynamic predictions 
for a monodisperse system and for a binary mixture.

\section{Molecular dynamics simulations for the one component system}
\label{sec:one}

\subsection{Introduction}

We have implemented molecular dynamics simulations with an 
event-driven algorithm
for $N$ spherical particles in a two-dimensional $L \times L$
box. Periodic boundary conditions are enforced in the $x$ direction,
while the energy loss due to collisions is compensated by an energy
injection by two walls situated at $y=0$ and $y=L$ 
(we consider the amplitude of motion of the walls to be small so that
their positions are considered as fixed \cite{Brey1}, which avoids 
the complication
of heat pulses propagating through the system \cite{McNamara}). 
We will refer to the $y$ direction as the ``vertical'' one,
although we are interested in regimes for which gravity can be 
neglected~\cite{Rouyer} (i.e. when the shaking is violent enough). 
The energy
can be injected in two ways: 
\begin{itemize}
\item by {\em thermal walls} which impose a given temperature of order 
$T_0$ \cite{Grossman}: when a 
particle collides with the wall, its new vertical (along $y$)
velocity is extracted at random from the probability distribution function
$v/\sqrt{T_0} \exp [ - v^2/(2 \sqrt{T_0}) ]$, whereas $v_x$ is unaffected.

\item by {\em vibrating walls}: for simplicity, we consider walls 
of infinite mass moving
in a sawtooth manner: all particles colliding with a wall find it with
the same velocity $v_0 >0$ at $y=0$, $-v_0$ at $y=L$. 
The particle-wall collisions
are considered elastic. A particle of velocity ${\bm v}$ with
$v_y < 0$ colliding with the bottom wall at $y=0$ (resp. $v_y > 0$ 
at the upper wall) sees its velocity change to ${\bm v}'$ according to
$v'_y=2v_0 - v_y$ (resp. $v'_y=-2v_0 - v_y$), whereas the $x$-component
is unaffected ($v'_x=v_x$).

\end{itemize}
In both cases, energy is injected in the vertical direction only,
and transferred to the other degrees of freedom through inter-particle
collisions.
The vibrating walls being the situation closer to the experimental one,
most of our results will be presented in this case, and the effect of 
injection modes will be briefly discussed.

In this section, we consider the monodisperse case: all particles
have the same mass $m(=1)$, diameter $\sigma$, restitution
coefficients $\alpha$ and $\alpha^t$. Most of the simulations are done
with $N=500$ particles, some with $N=1000$ particles (low enough 
to avoid clustering or inelastic collapse).
For our two-dimensional system, the local packing fraction at height 
$y$, where the local density is $\rho(y)$, is defined as 
$\eta(y) = \pi \rho(y) \sigma^2/4$. The global (mean) packing
fraction is denoted $\eta_0$: $\eta_0 = \int_0^L \eta(y) dy/L$. 

Starting from a random configuration of the particles (with the constraint
of no overlap), we let the system
evolve until a steady state is reached. Data on density and temperature
profiles as well as on velocity distributions are monitored as a time
averages; the various quantities are averaged along the $x$ direction since
the system remains homogeneous in this direction.

\subsection{Density and temperature profiles}

\begin{center}
\begin{figure}[ht]
\epsfig{figure=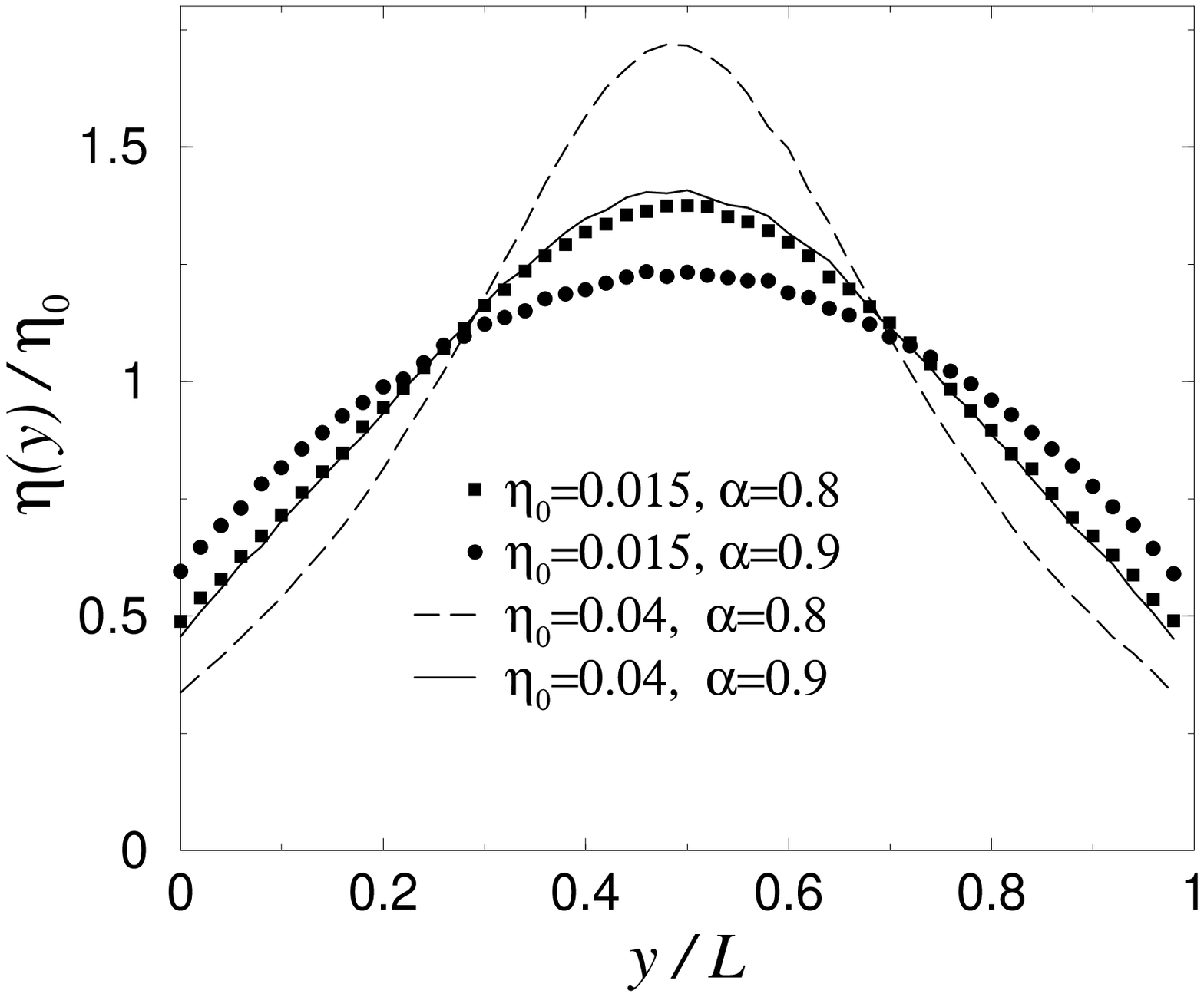,width=8cm,angle=0}
\caption{Density profiles for two normal inelasticities and two densities. 
In all cases, the number of particles is $N=500$. The symbols correspond
to the smallest density (the mean packing fraction, averaged over the whole
system is $\eta_0=0.015$) and the lines are for a higher density 
($\eta_0 = 0.04$). The ratio $\eta(y)/\eta_0$ is also the ratio 
$\rho(y)/\rho_0$
of local density normalized by the mean one.}
\label{fig:dens}
\end{figure}
\end{center}

\begin{center}
\begin{figure}[htb]
\epsfig{figure=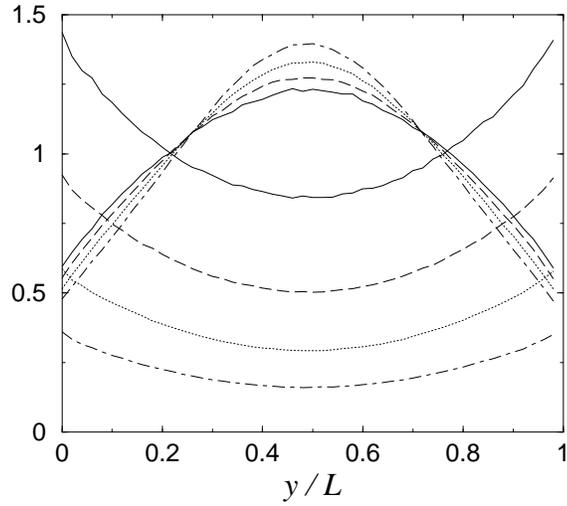,width=8cm,angle=0}
\caption{Density profiles $\eta(y)/\eta_0$ 
(upward curves) and temperature profiles 
(downward curves) for 
a given normal restitution coefficient $\alpha=0.9$, and different
tangential restitutions 
($N=500$ particles, mean packing fraction $\eta_0 = 0.015$). 
The temperature is the total one (including horizontal and vertical
degrees of freedom); it is expressed in arbitrary units but all curves
correspond to the same velocity of the vibrating piston. From 
top to bottom for the temperature $T(y)$ and from bottom to top
for the density, the curves correspond respectively to 
$\alpha^t=-1$, $\alpha^t=-0.8$, $\alpha^t=-0.5$ and $\alpha^t=0.2$.
}
\label{fig:densrot}
\end{figure}
\end{center}

The first observations concern the density and temperature profiles:
Figs.~\ref{fig:dens} and \ref{fig:densrot} show that the density is lower
near the walls, where the temperature is higher as expected since
energy is injected at the walls and dissipated in the bulk of the system
\cite{Helal}. 
The profiles are qualitatively similar for
thermal or vibrating boundaries.
Moreover, the whole temperature profile
is proportional to the temperature $T_0$ imposed by a thermal wall
or to the square of the velocity $v_0$ of the vibrating boundary, 
while a change
in $T_0$ or $v_0$ does not change the density profile (not shown). As the mean
density increases or $\alpha$ decreases, the profiles get more heterogeneous;
as $\alpha^t$ is increased, more energy is transferred to rotational
degrees of freedom, 
so that the temperature decreases, while the density profiles become
slightly more peaked (Fig.~\ref{fig:densrot}).

\begin{center}
\begin{figure}[htb]
\epsfig{figure=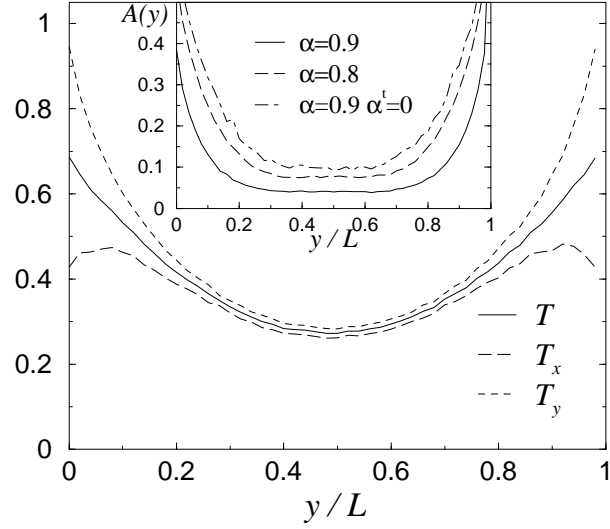,width=8cm,angle=0}
\caption{Temperature profile for $\alpha=0.9$ and $\eta_0 = 4\%$.
The horizontal $T_x$, vertical $T_y$ and total temperature $T=(T_x+T_y)/2$
are shown. Inset: anisotropy factor $A=(T_y-T_x)/(2T)$ as a function of 
height.}
\label{fig:aniso}
\end{figure}
\end{center}

Fig.~\ref{fig:aniso} clearly shows another feature 
resulting from the energy injection into the vertical direction: the 
temperature is anisotropic, i.e. 
$\langle v_x^2 \rangle \ne \langle v_y^2 \rangle$, with $T_y > T > T_x$.
The anisotropy $A(y) = (T_y - T_x)/(2T)$ is larger at the boundaries, where
energy is fed into the vertical direction, decreases due
to inter-particle collisions, and reaches a
plateau in the middle of the slab. The plateau value 
decreases for increasing number of particles or increasing densities
(not shown), as in experiments~\cite{Rouyercont}; the global
anisotropy profile and the plateau values are 
comparable to experimental values~\cite{Rouyercont}.

\subsection{Equation of state and hydrodynamics}

The equation of state derived in section \ref{sec:eos}
reduces, in the case
of a two dimensional one-component homogeneous system, to the relation
\begin{equation}
P= \rho T \left[ 1 + (1+\alpha)\,\eta\, \chi \right] ,
\label{eq:Ponecomp}
\end{equation}
where $\chi$, the pair correlation function at contact, depends on
the packing fraction $\eta$. We will use the 
form $\chi = (1-7\eta/16)/(1-\eta)^2$, which has been shown to be accurate 
for elastic hard disk liquids \cite{Henderson}. 

\begin{center}
\begin{figure}[htb]
\epsfig{figure=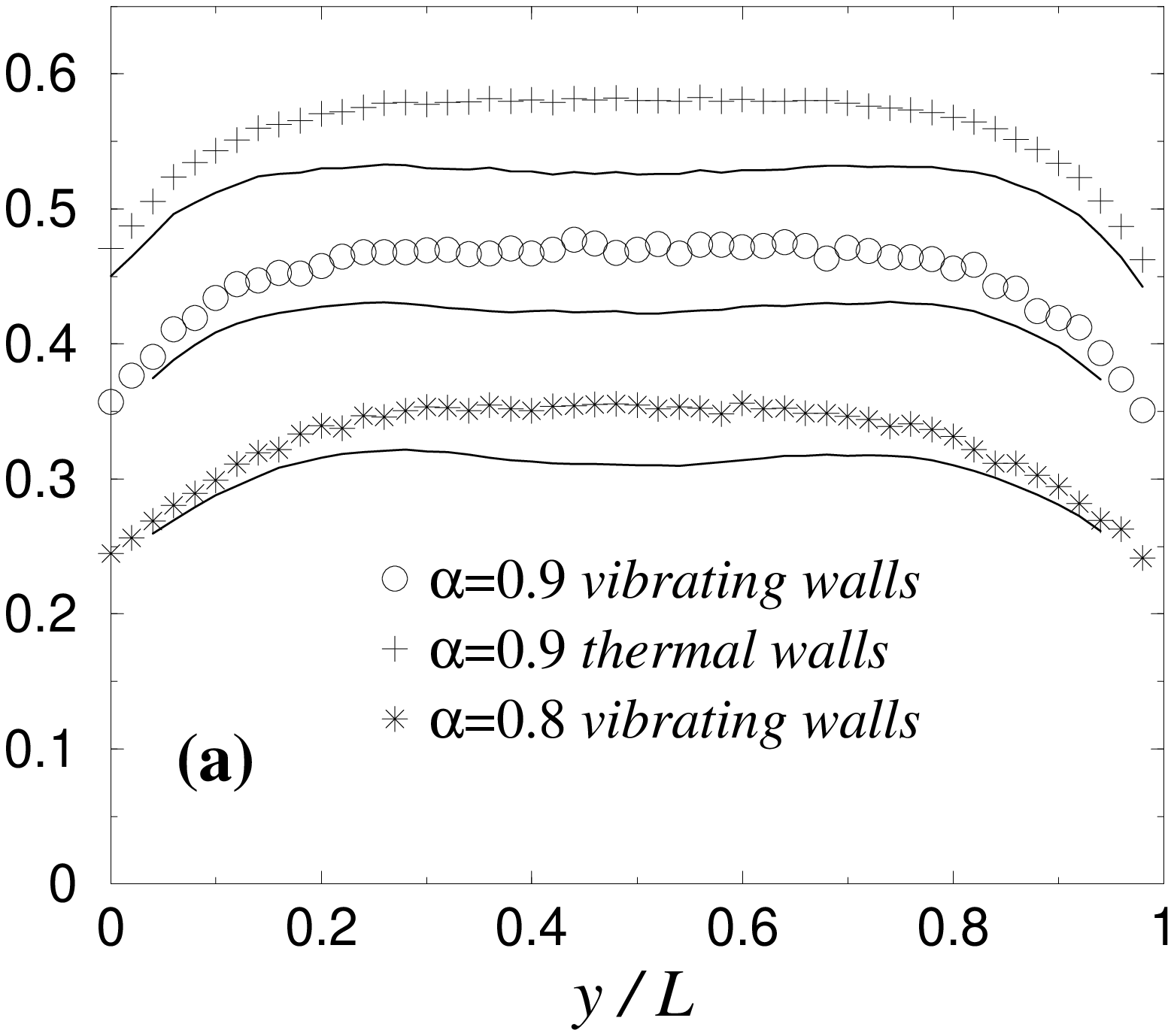,width=8cm,angle=0}
\epsfig{figure=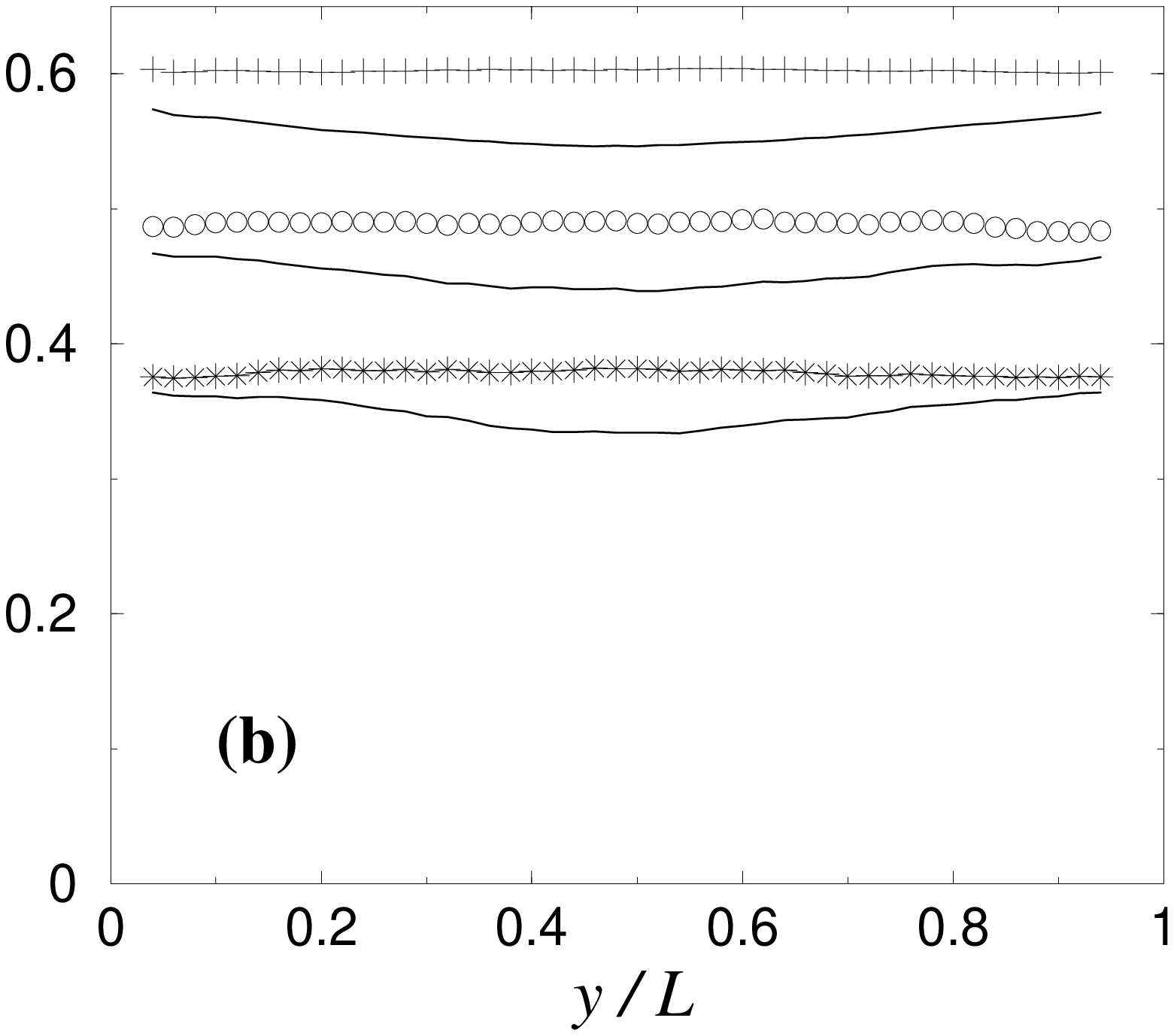,width=8cm,angle=0}
\caption{Pressure given by the equation of state (\protect\ref{eq:Ponecomp}). 
(a) The symbols correspond to 
$P=\rho(y) T(y) [1+(1+\alpha)\eta(y) \chi(y)]$ (see text), where $T$ is the 
total temperature. The lines immediately below a given set of symbols show the
ideal gas contribution $\rho(y) T(y)$ only. For the three situations 
investigated, the mean density is the same ($\eta_0 = 0.04$). \\
(b) Same figure with
the vertical temperature $T_y$ instead of $T$ inserted in the equation
of state, yielding therefore the $yy$ component of the pressure tensor.}
\label{fig:znT}
\end{figure}
\end{center}

The hydrodynamic equations (see appendix \ref{app:hydro}, and \cite{Brey1})
lead to $\partial_y P = 0$ in the absence of a flow field. 
We check in Fig.~\ref{fig:znT}(a) the constancy of $P$ with $y$
by plotting the ideal gas contribution
$\rho(y) T(y)$ (lines) and $P(y)$ given by Eq. (\ref{eq:Ponecomp}) 
(i.e. ideal gas contribution plus Enskog correction).
While, at small enough densities (not shown), 
$\rho(y) T(y)$ is constant in the bulk
($y \in [0.2L, 0.8L]$), the Enskog correction is necessary for the densities
used in Fig.~\ref{fig:znT} (note that the density can be quite larger
in the middle of the system than at the boundaries). We also note that
the inelasticity term $(1+\alpha)$ is relevant [the profiles of
$\rho T \left( 1 + 2 \eta \chi \right)$, not shown,
do not display a uniform shape with $y$]. In all cases,
boundary layers ($y < 0.2L$ and $y>0.8L$) are observed~\cite{Brey1} in which
the pressure decreases. This discrepancy can be related
to the anisotropy described in the previous subsection (pressure and
temperature are most anisotropic near the walls). 

The comparison with hydrodynamics may be improved as follows.
The pressure tensor ${\bm P}$ is diagonal in the present
no flow situation, but has different $xx$ and $yy$ components, 
and the homogeneity along the $x$ direction implies that the
condition of vanishing flow field $\nabla \cdot {\bm P}=0$ 
reduces to $\partial_y P_{yy} = 0$.
We therefore check in Fig.~\ref{fig:znT}(b) that the $yy$ component
of the pressure tensor, given by the equation of state (\ref{eq:Ponecomp})
with the total temperature $T=(T_x+T_y)/2$ replaced by its vertical component
$T_y$, is uniform in the whole system. With Enskog correction,
the corresponding profiles are remarkably flat. 
This result could be tested in experimental situations in which
both $T_x$ and $T_y$ are measured. Such an analysis validates
both the hydrodynamic picture and 
the equation of state proposed by automatically sampling several
densities in a single run.

\begin{center}
\begin{figure}[ht]
\epsfig{figure=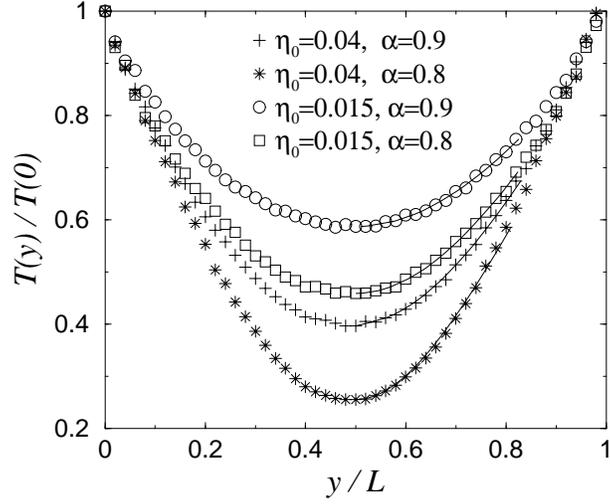,width=8cm,angle=0}
\caption{Fits of the temperature profiles measured in MD with the analytical
expression (\ref{eq:fit}). The fits are shown with continuous curves while
the symbols stand for the MD measures.
For clarity the fits are restricted to heights
$L/2 \leq y \leq 0.8 L$.  }
\label{fig:temp}
\end{figure}
\end{center}

At low densities, assuming the ideal gas equation of
state to hold, the hydrodynamic study of Ref.~\cite{Brey1}
(recalled in appendix~\ref{app:hydro}), leads to the following analytical
prediction for the temperature profile:
\begin{eqnarray}
\frac{y}{L}&=& \frac{\xi + \sinh \xi \cosh (\xi_m - \xi)}
{\xi_m + \sinh \xi_m} \nonumber \\
\xi &=&  \frac{\xi_m}{2} \pm \cosh^{-1} \left( \sqrt{\frac{T}{T_0}} 
\cosh \frac{\xi_m}{2} \right),
\label{eq:fit}
\end{eqnarray}
where $T_0$ is the temperature at the boundaries and $\xi_m$ is proportional
to the total number of particles.
The corresponding  fits of the temperature profiles are shown in 
Fig.~\ref{fig:temp}; a good
agreement is obtained, especially at lower densities as expected
[since the ideal gas equation of state is a crucial ingredient
in the derivation of Eqs. (\ref{eq:fit})].
We use one fitting parameter $\xi_m$ to obtain $T/T_0$. 
Fig.~\ref{fig:znT} showed that consideration of the ``vertical''
pressure $P_{yy}$ led to a better agreement with hydrodynamic
predictions than the total $P_{xx}+P_{yy}$. A similar conclusion is 
incorrect for the temperature profiles:
the transport equation for the temperature is scalar [see Eq. (\ref{eq:c2})]
and Eqs. (\ref{eq:fit}) hold for
the total $T$, not for the vertical $T_y$.

\subsection{Velocity distributions}

\begin{center}
\begin{figure}[htb]
\epsfig{figure=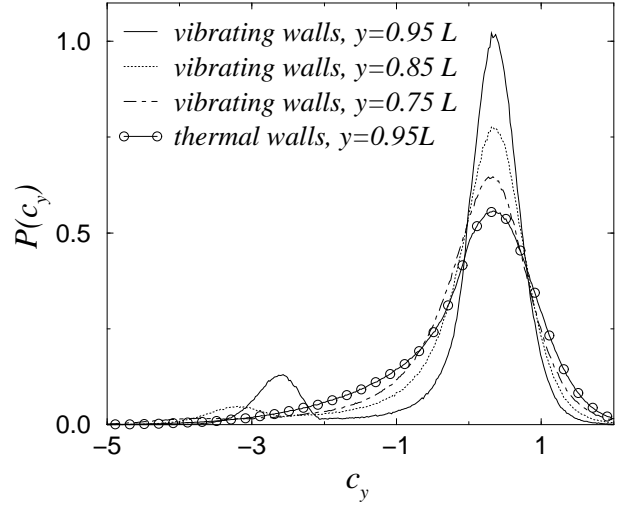,width=8cm,angle=0}
\caption{Probability distribution function (p.d.f.) of the 
vertical velocity component 
$c_y = v_y/\sqrt{T_y}$, for different heights. By definition, 
$\langle c_y^2\rangle = 1$ whatever the altitude $y$. Here, 
$\eta_0 = 0.04$, $N=500$, $\alpha=0.9$ and $\alpha^t=0$. }
\label{fig:pvy}
\end{figure}
\end{center}

Because of the energy injection through the walls, the velocity
distributions are anisotropic, and a priori depend on the distance to the
walls. The vertical velocity distribution also depends on the nature
of the walls as shown in Fig.~\ref{fig:pvy}. A smooth distribution
is obtained in the vicinity of a thermal wall, while the incoming and
out-coming particles yield two separated peaks for vibrating
walls (see also~\cite{Brey1}). 

\begin{center}
\begin{figure}[htb]
\epsfig{figure=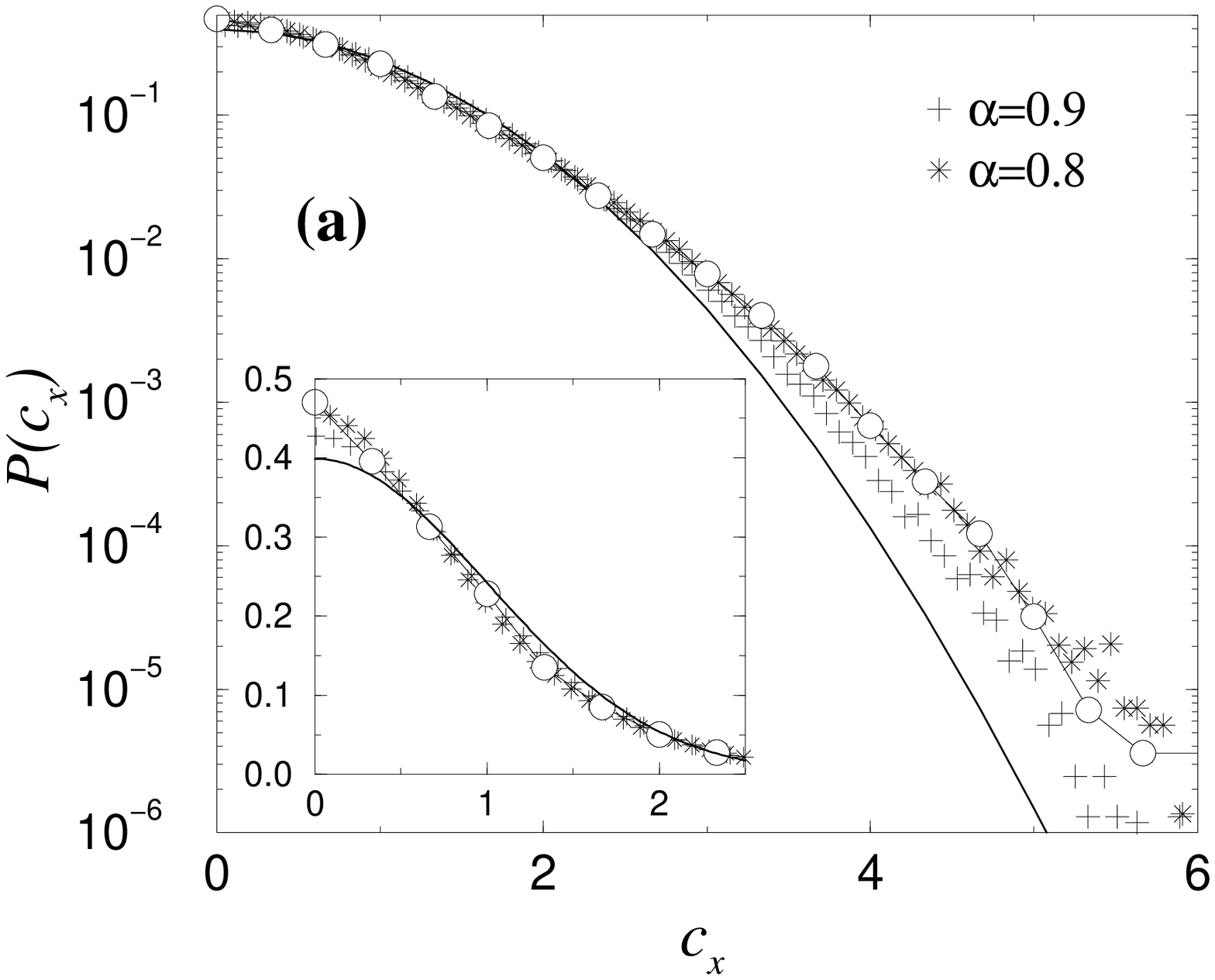,width=8cm,angle=0}
\epsfig{figure=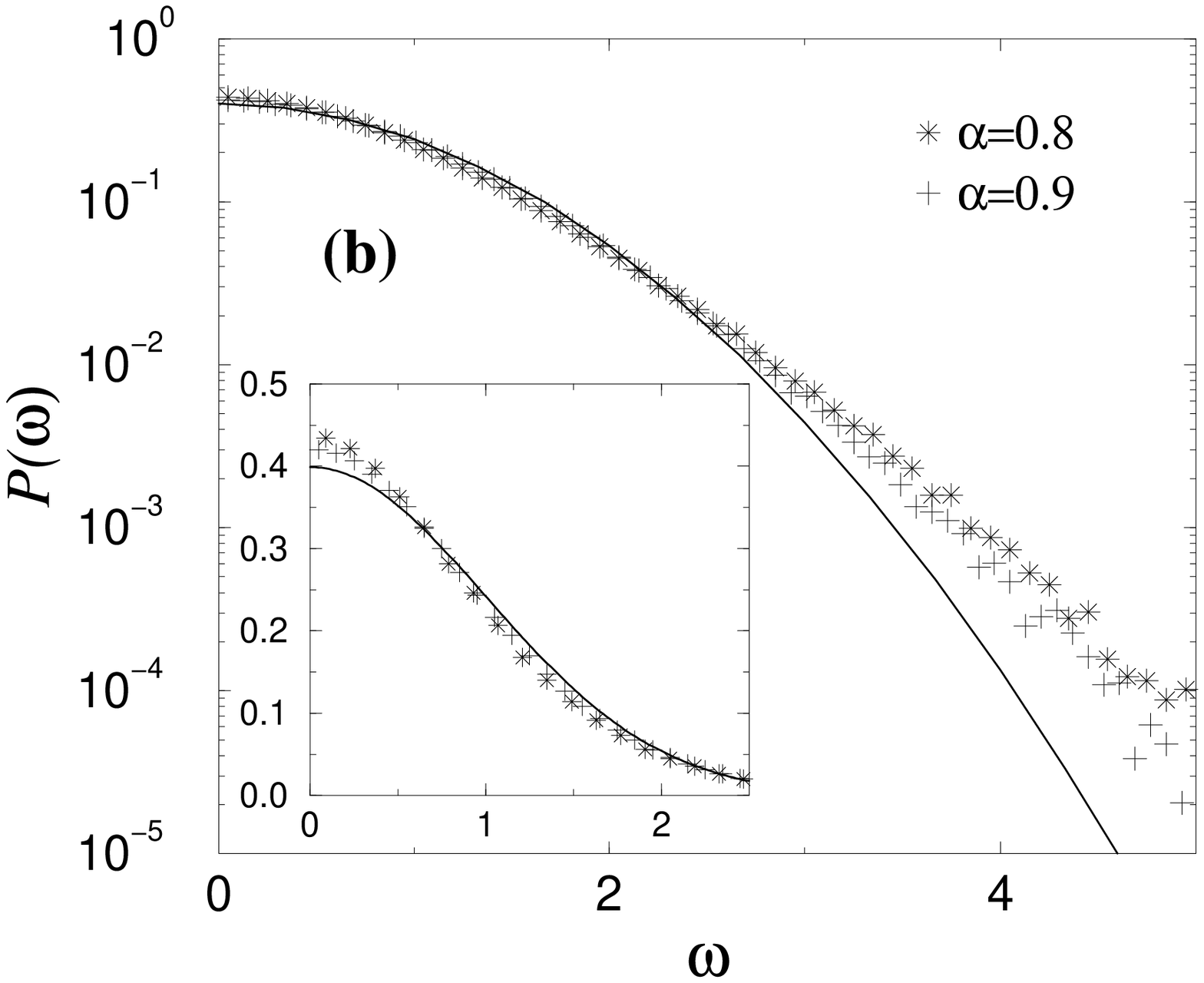,width=8cm,angle=0}
\caption{(a): Probability distribution function of the 
rescaled horizontal velocity component $c_x = v_x/\sqrt{T_x}$,
on a linear-log plot.
Here $\eta_0 = 0.015$, $N=500$, $\alpha=0.9$ (pluses) and $0.8$ (stars),
and $\alpha^t=0$.
The solid line is the Gaussian with unit variance, 
the circles correspond to experimental
data \protect\cite{Rouyer,Feitosa} for steel beads. \\
(b): Probability distribution function of the angular velocities
for the same parameters.
}
\label{fig:pvx}
\end{figure}
\end{center}

On the other hand, the {\em rescaled} horizontal velocity distribution 
$P(c_x)$ (with $c_x=v_x/\sqrt{T_x}$)
is remarkably independent of the distance from the walls
(outside the boundary layers), even if the temperature changes with $y$.
Fig.~\ref{fig:pvx} shows clearly non-Gaussian features similar to the
experimentally observed ones~\cite{Rouyer,losert,Aranson}, with in particular
overpopulated both small-velocity and high-velocity regions. A
slight dependence on the parameters is obtained: 
$P(c_x)$ broadens if the inelasticity increases
(i.e. if $\alpha$ decreases), if $\alpha^t$ increases, or if $\eta_0$ 
or $N$ increase. Experimentally, the dependence on density or material
properties is weak and difficult to measure
\cite{Rouyer} but seems to exist, in particular
as far as $N$ is varied \cite{Feitosa}.
The angular velocity distributions, also displayed in Fig.~\ref{fig:pvx},
share a similar non-Gaussian character and the same dependence with the
parameters.

As density or inelasticity are further increased, clustering
phenomena may occur, leading to heterogeneities along the $x$ direction,
with the coexistence of colder, denser regions with hotter, less
dense ones. The average over the $x$ direction then leads to artificially
broad $P(c_x)$.

Finally, as a general rule, thermal walls
lead to slightly broader velocity distributions than vibrating walls.

\subsection{Effective restitution coefficients}

We now turn to the study of the effective inelasticities introduced
in the context of a Random Restitution Coefficient model (RRC)
\cite{RRC1,RRC2}: even if the restitution coefficient
$\alpha$ is constant, the energy is injected in the vertical
direction and transferred to other degrees of freedom through collisions, so
that the {\em effective} restitution coefficient for collisions
projected onto the $x$ direction,
\begin{equation}
\alpha_{1d} = \frac{v_{12,x}'}{v_{12,x}} \ ,
\end{equation}
may be either smaller or larger than $1$. This leads to the definition
of a one-dimensional effective model with a restitution
coefficient taken at random from a given distribution
at each collision~\cite{RRC1,RRC2}.

\begin{center}
\begin{figure}[ht]
\epsfig{figure=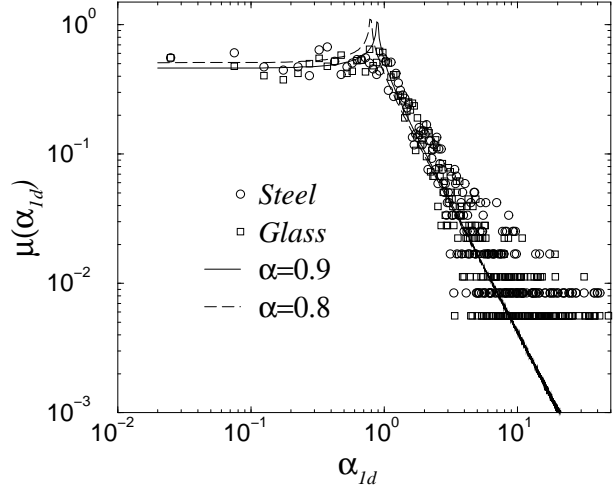,width=8cm,angle=0}
\caption{Probability distribution function of effective one-dimensional
restitution coefficients $\alpha_{1d}$. The MD results are compared to the 
experimental measures of Feitosa and Menon \protect\cite{Feitosa}
on steel and glass samples (for which the nominal restitution coefficient
may be considered close to $0.9$). 
}
\label{fig:hista1d}
\end{figure}
\end{center}

Values of $\alpha_{1d}$ have been experimentally measured~\cite{Feitosa,RRC2}
and shown to display a broad probability distribution 
$\mu(\alpha_{1d})$ very similar for various materials and densities. 
We have measured $\alpha_{1d}$ for many collisions and thus obtained
its distribution, displayed in Fig.~\ref{fig:hista1d} together
with experimental data for steel and glass beads. A remarkable
agreement is found. Our study shows that $\mu(\alpha_{1d})$ display
a $\alpha_{1d}^{-2}$ tail for $\alpha_{1d} > 1$, irrespective of
$\alpha$, $\alpha^t$ and density. 

\begin{center}
\begin{figure}[ht]
\epsfig{figure=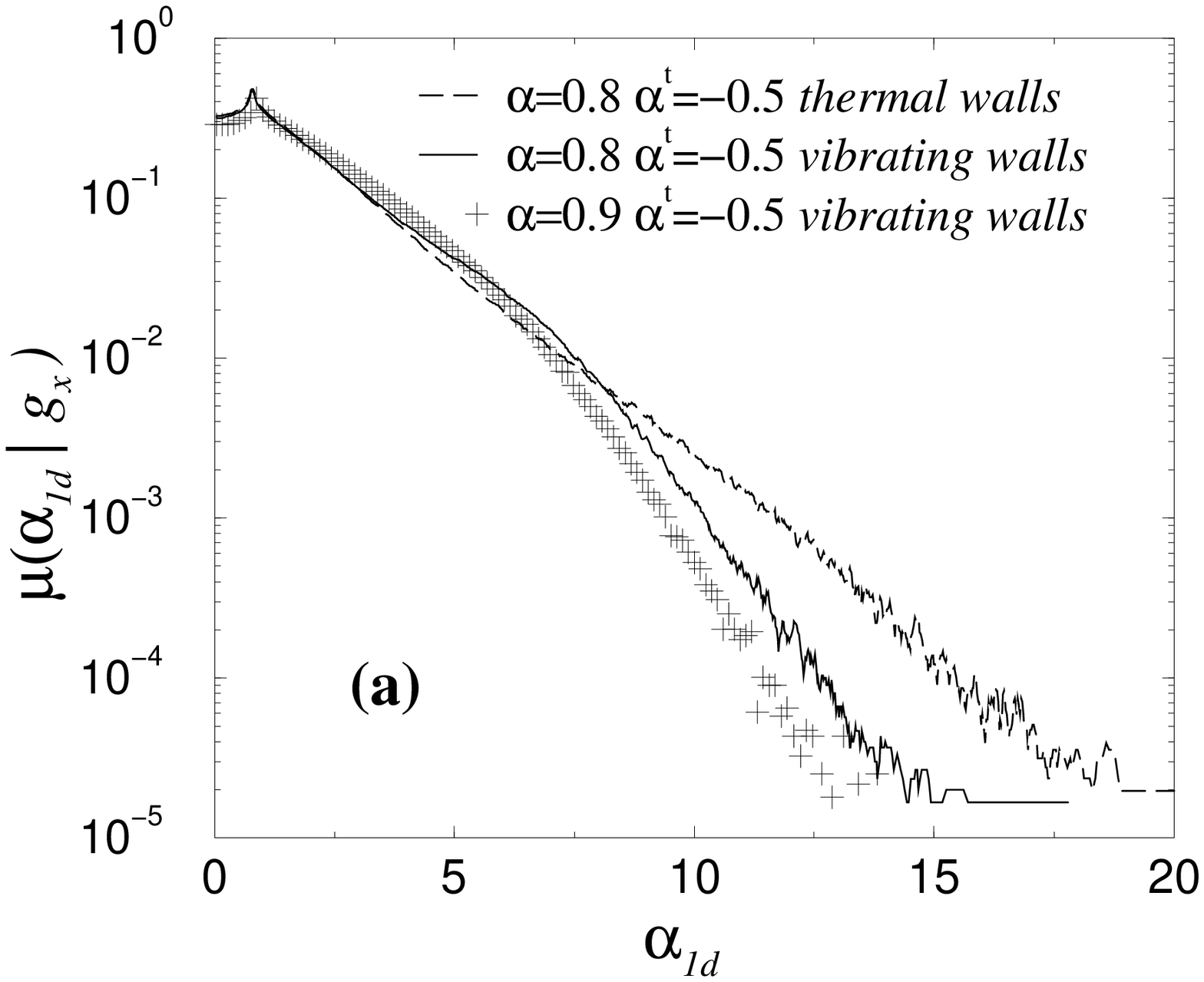,width=8cm,angle=0}
\epsfig{figure=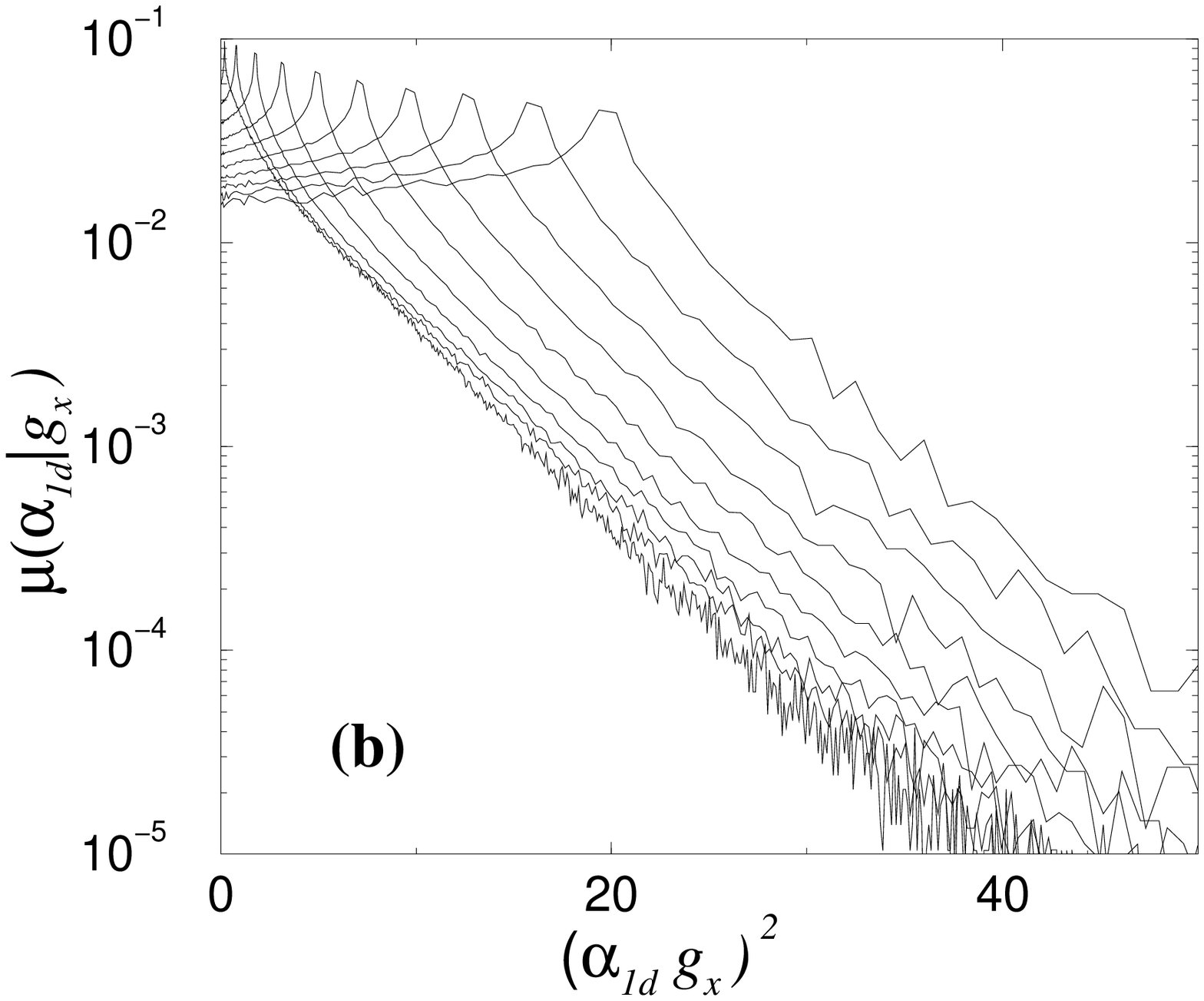,width=8cm,angle=0}
\caption{(a): Conditional p.d.f. of $\alpha_{1d}$ 
for a given value $g_x$ of order unity. Note the different
shapes for thermal and vibrating walls.\\
(b): Same, as a function of
$(\alpha_{1d} g_x)^2$ (and
$g_x= 0.2, 0.5, 1., 1.5, 2., 3., 4., 5.$) for vibrated walls with
$\alpha=0.9$, $\alpha^t=0$ and
$\eta_0 =0.015$.
}
\label{fig:hista1dgx}
\end{figure}
\end{center}

The importance of the correlations
between $\alpha_{1d}$ and the relative velocity 
${\bm g}={\bm v}_{12}/\sqrt{2T}$ of the colliding particles has been
emphasized in~\cite{RRC2} and is revealed by the computation 
of $\mu(\alpha_{1d}|g_x)$, the distribution of $\alpha_{1d}$
conditioned by a given value of $g_x$; 
although no precise experimental determination of 
the conditional $\mu(\alpha_{1d}|g_x)$ could be achieved in \cite{RRC2}, 
strong evidences for a sharp cut-off $\propto 1/g_x$ at large values of
$\alpha_{1d}$ were provided and the form
$\mu(\alpha_{1d}|g_x)\propto \exp\left(- (\alpha_{1d}g_x)^2/R \right)$ at large
$\alpha_{1d}$ has been proposed.
The conditional $\mu(\alpha_{1d}|g_x)$ obtained in the present MD
simulations confirm the above picture; they are displayed in
Fig.~\ref{fig:hista1dgx} and show an
$\exp\left[- (\alpha_{1d}g_x)^2/R \right]$ decrease for 
the case of vibrating walls (closer to the experimental situation),
and a broader form
$\exp\left[- (\alpha_{1d}g_x)/R' \right]$ for thermal walls.
Moreover, although $\mu(\alpha_{1d})$ is not sensitive to the various
parameters, the cut-off $R$ increases [i.e. leads to broader
$\mu(\alpha_{1d}|g_x)$] if $\alpha$ decreases, and if $\alpha^t$
or $\eta_0$ increase.

These findings, together with the evolution of the velocity
distributions $P(c_x)$ with the parameters, is in complete agreement
with the one-dimensional effective RRC model put forward in~\cite{RRC2},
for which broader conditional distributions $\mu(\alpha_{1d}|g_x)$
are linked to broader $P(c_x)$ (at large $c_x$, compared to the Gaussian).

\begin{center}
\begin{figure}[ht]
\epsfig{figure=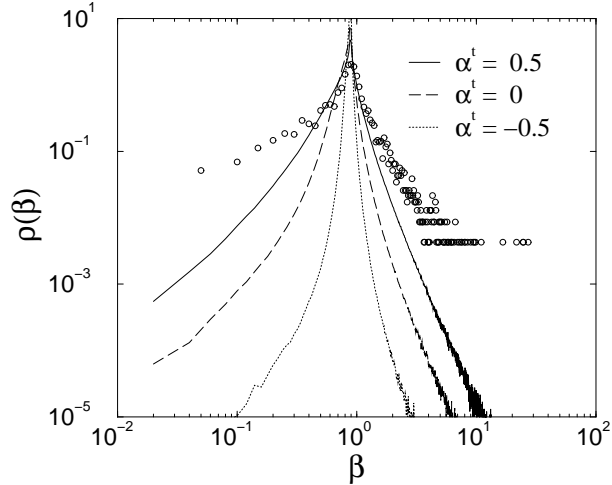,width=8cm,angle=0}
\caption{Probability distribution function of energy restitution
coefficients $\beta$. Various tangential restitution coefficients 
$\alpha^t$ are
considered for $\alpha=0.9$ and $\eta_0=1.5\%$. 
The circles represent the experimental data 
for steel grains \protect\cite{Feitosa}}
\label{fig:histb}
\end{figure}
\end{center}

Finally, the {\em energy} restitution coefficient
\begin{equation}
\beta = \frac{ v_{12}'}{v_{12}}
\end{equation}
may also be viewed as a random variable that can take values larger than unity
due to energy transfers between rotational and 
translational degrees of freedom~\cite{Feitosa,RRC2}. 
Fig.~\ref{fig:histb} displays 
the p.d.f. $\rho(\beta)$ obtained in the MD simulations for
various values of $\alpha^t$, together with the experimental data
of~\cite{Feitosa,RRC2} for steel beads. $\rho(\beta)$ becomes
wider as $\alpha^t$ is increased, but the experimental distribution
is broader, which may be traced back to the fact that in the experiments
mentioned above,
the beads can rotate in three dimensions whereas our simulations
are limited to 2D rotations.

\section{Molecular dynamics simulations for the binary mixture}
\label{sec:binary}

In this section, we investigate the properties of vibrated binary mixtures;
such systems have recently attracted much attention, 
both on the experimental \cite{losert,Wildman,menon} and theoretical side
\cite{Duparcmeur,garzo,GarzoDufty,Clelland,Dahl,MontaneroShear,%
MontaneroHCS,Equipart,puglisi2,Pagnani,biben}.
In particular the breakdown of energy equipartition between the two
constituents of the mixture has been thoroughly investigated.

The main difference with previous studies consists here in the
realistic character of both MD simulations (as opposed to Monte Carlo methods)
and 
the energy injection mechanism at the boundaries;
the set-up is the same as in the previous section, with however
two types of particles, with
masses $m_1$, $m_2$, sizes $\sigma_1$, $\sigma_2$. The
three normal restitution coefficients (corresponding to the 
three possible types of collisions) are 
$\alpha_{11}$, $\alpha_{12}=\alpha_{21}$, $\alpha_{22}$.
In the context of a forcing mechanism through a random external 
force \cite{Pre1,Pre2}, it has been shown that the influence of size ratio
on the temperature ratio measuring the energy non-equipartition
was rather weak \cite{Equipart} compared to that of inelasticity parameters
or mass ratio. We shall consequently limit our study to identical 
sizes $\sigma_1=\sigma_2$ in two dimensions, which corresponds to 
the experimental situation we will refer to \cite{menon,Feitosa}.
For simplicity,
the tangential restitution coefficients $\alpha_{ij}^t$ are also taken
equal.

As in the monodisperse case, we measure density and temperature profiles, 
velocity distributions as well as the temperature ratios
$\gamma (y) =T_2(y)/T_1(y)$, $\gamma_x (y) =T_{2,x}(y)/T_{1,x}(y)$,
$\gamma_y (y) =T_{2,y}(y)/T_{1,y}(y)$.
Some comparison with experimental data~\cite{menon,Feitosa} will be proposed
whenever possible.

\subsection{Equation of state}

\begin{center}
\begin{figure}[ht]
\epsfig{figure=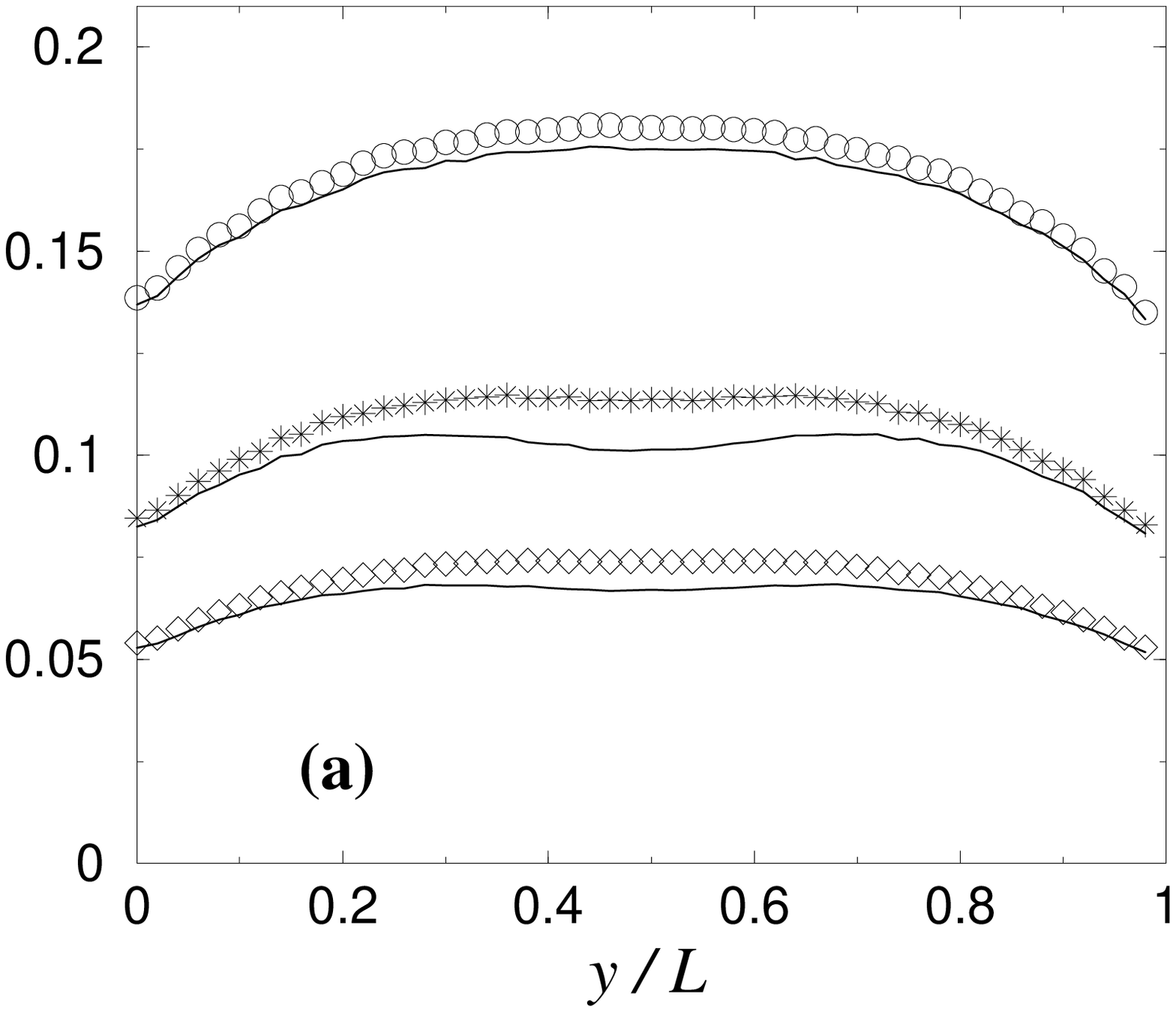,width=8cm,angle=0}
\epsfig{figure=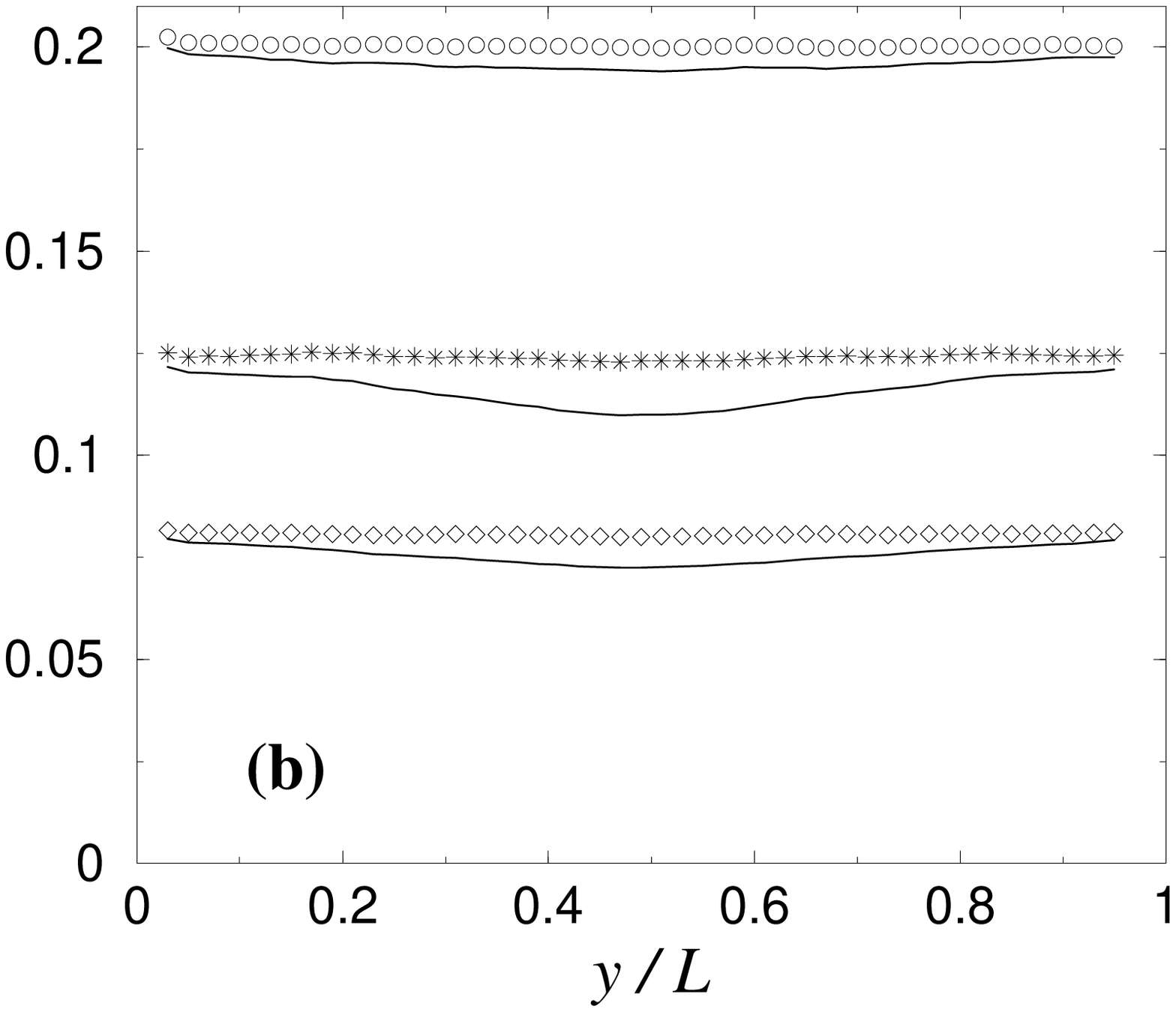,width=8cm,angle=0}
\caption{(a): The symbols show the pressure calculated
from the complete equation of state for a binary mixture
(\protect\ref{eq:multi}) including Enskog correction, while the lines
immediately below display the ideal gas contribution 
$\rho_1(y) T_1(y) + \rho_2(y)T_2(y)$ to the pressure. 
The three sets of curves correspond
to: {\em upper set} 
$\eta_0 = 0.015$, $\alpha_{11} = 0.9$, $\alpha_{12} = 0.8$, 
$\alpha_{22} = 0.7$,
$m_1=5m_2$;
{\em middle set} 
$\eta_0 = 0.04$, $\alpha_{11} = 0.9$, $\alpha_{12} = 0.8$, $\alpha_{22} = 0.7$,
$m_1=3m_2$;
{\em lower set} 
$\eta_0 = 0.04$, $\alpha_{11} = 0.7$, $\alpha_{12} = 0.8$, $\alpha_{22} = 0.9$,
$m_1=3m_2$.\\
(b): same curves, where the temperatures are the vertical ones 
$T_{i,y}$ instead of the total $T_i = (T_{i,x}+T_{i,y})/2$, yielding
therefore the $yy$ component of the pressure tensor.
}
\label{fig:binznT}
\end{figure}
\end{center}

We first test the equation of state (\ref{eq:eos}) in 
Fig.~\ref{fig:binznT}. As in the monodisperse case, the Enskog
correction is clearly relevant, even at low global densities, since
the density profiles reach relatively high values for $y\simeq L/2$. 
It is however sufficient to truncate the equation of state at second
virial order, which amounts to take the low density limiting value
$\chi_{ij}=1$ for the pair correlation functions at contact:
\be
P \,\simeq \, \rho_1 T_1 + \rho_2 T_2 +  
\frac{\pi \sigma^2}{2(m_1+m_2)}\,
\left[(1+\alpha_{11}) \rho_1^2 m_2 T_1 + (1+\alpha_{12}) \rho_1 \rho_2 
(m_1 T_2+m_2 T_1) + (1+\alpha_{22}) \rho_2^2 m_1 T_2
\right].
\label{eq:multi}
\ee
Moreover, the boundary
layer in which the anisotropy is strong is still apparent if the global
temperatures $T_1$ and $T_2$ are used, while use of the vertical ones $T_{1,y}$
and $T_{2,y}$,
suggested by the anisotropy of temperatures and pressure
as in the monodisperse case,
leads to a uniform $yy$ component of the pressure tensor
in the whole system. The functional dependence of pressure
upon densities is therefore accurately reproduced by the
equation of state (\ref{eq:multi}).

\begin{center}
\begin{figure}[ht]
\epsfig{figure=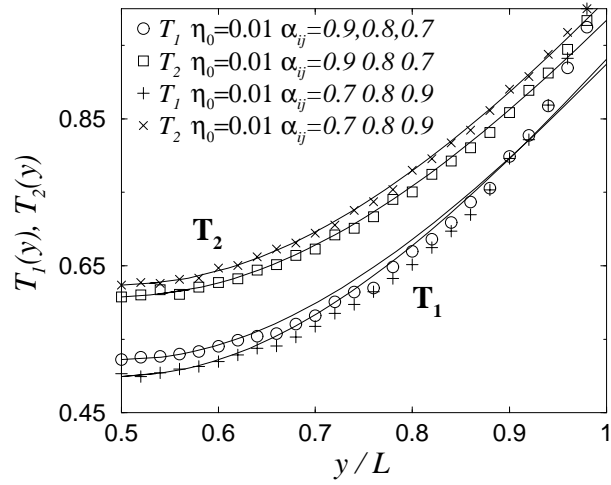,width=8cm,angle=0}
\caption{Temperature profiles for an equimolar granular mixture, driven by
vibrating walls. The symbols show the MD
measures, and the lines are fits  to the analytical expression derived 
for the single component case. In all cases, the particle 1 (the 
heaviest)
has mass $m_1=3m_2$; its temperature $T_1$ corresponds to the two lower sets.}
\label{fig:fitstemp}
\end{figure}
\end{center}

Although we have not extended the hydrodynamic approach of Brey {\it et al.}
\cite{Brey1}
to binary mixtures (it would be possible making use of the
Navier-Stokes like equations derived in \cite{GarzoDufty}
where only the overall temperature associated with both species 
serves as a hydrodynamic field, but where the transport coefficients
explicitly depend on temperature ratio), 
we see in Fig.~\ref{fig:fitstemp} that the temperature
profiles can be fitted, at low density, by the form (\ref{eq:fit}). 
We emphasize that there is no fundamental reason for the agreement.
The quality of the fit is much better for the less massive particles
whose density is more homogeneous across the system (see next subsection).
For simplicity, we have used the short hand notation 
$\alpha_{ij} = 0.7; 0.8; 0.9$
for the situation where  
$\alpha_{11}=0.7$, $\alpha_{12}=0.8$ and $\alpha_{22}=0.9$.

\subsection{Non equipartition of translational kinetic energy}

\begin{center}
\begin{figure}[ht]
\epsfig{figure=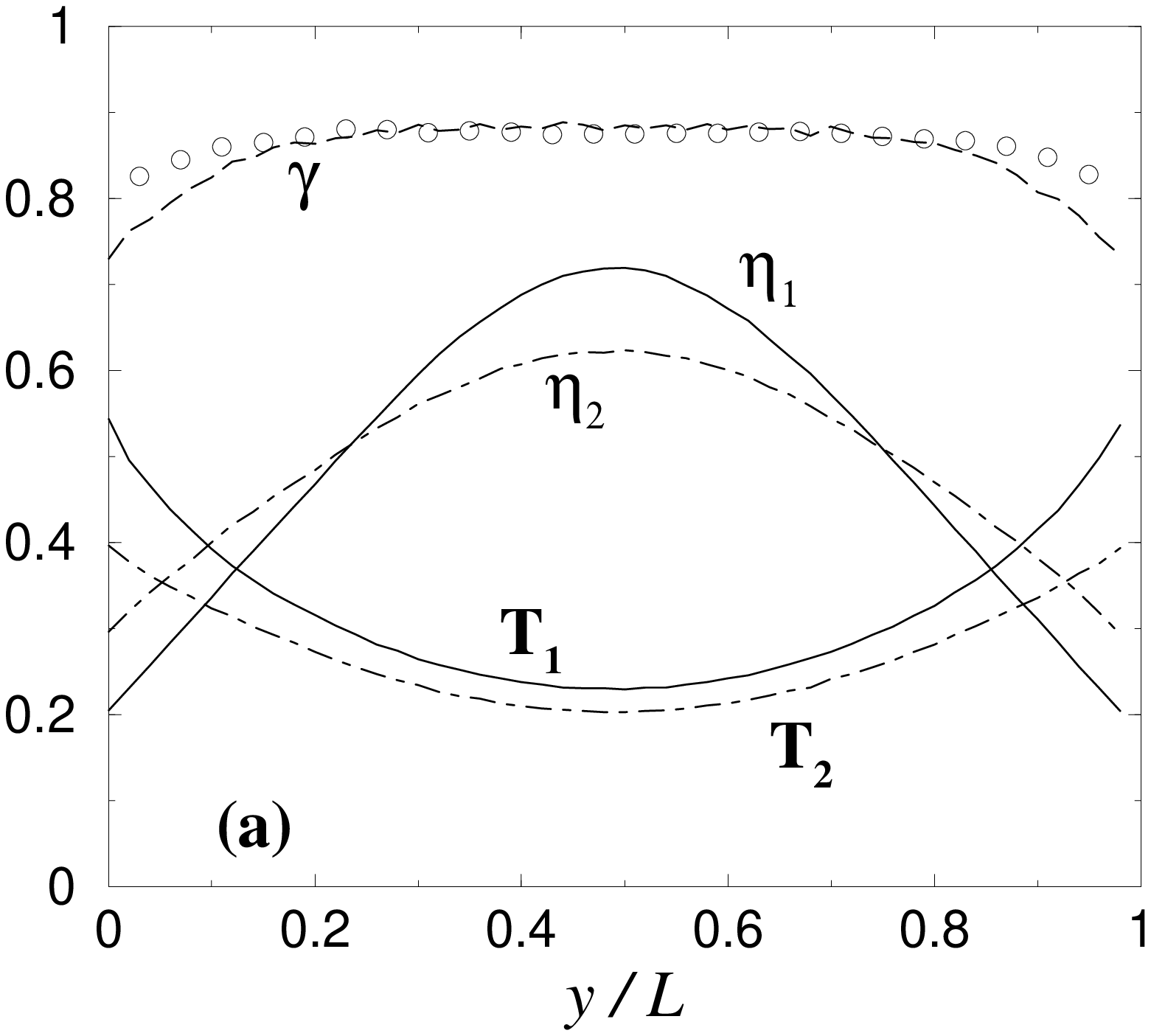,width=8cm,angle=0}
\epsfig{figure=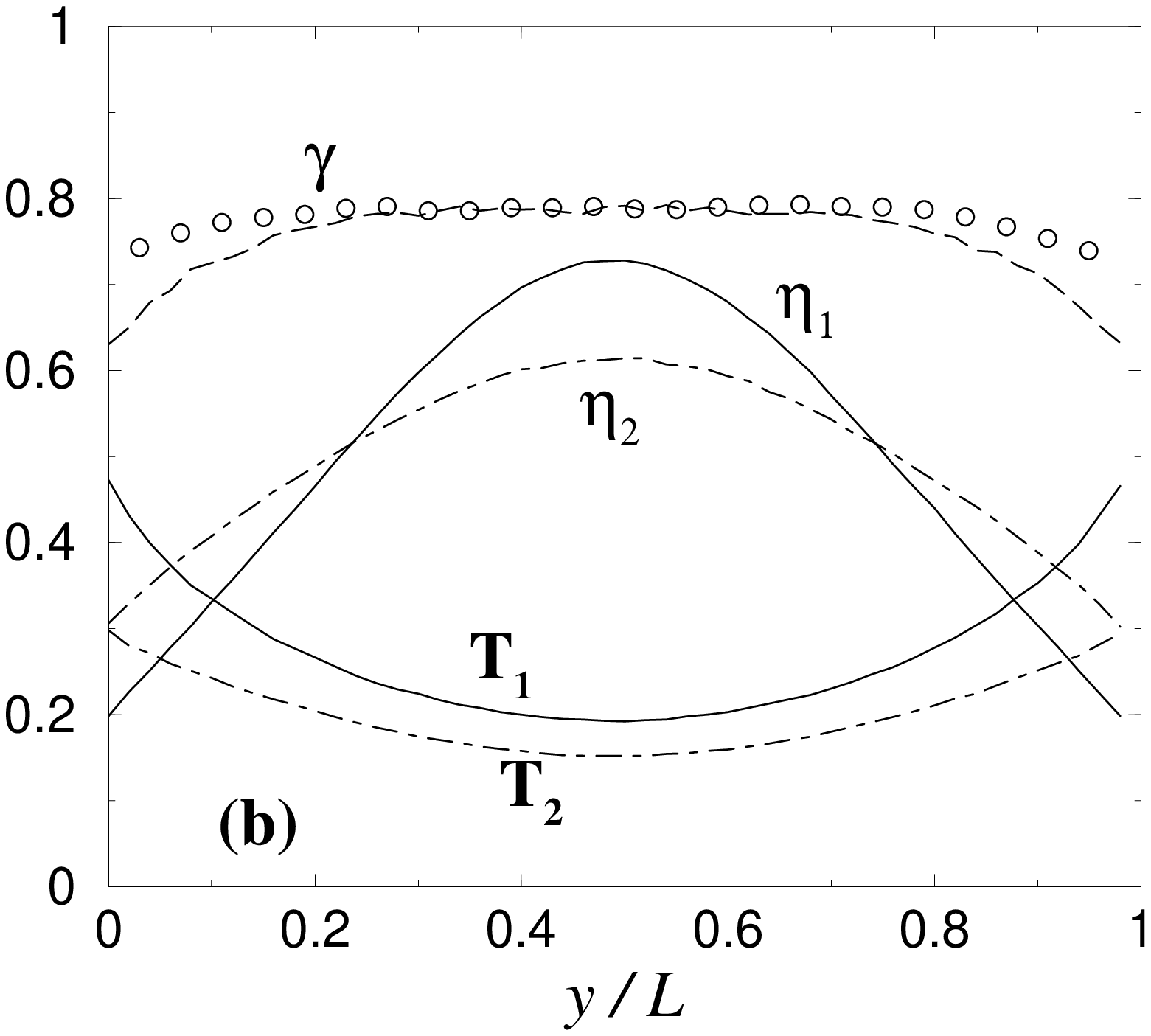,width=8cm,angle=0}
\caption{(a): Vertical profiles for a binary mixture with $m_1=3m_2$, 
$\eta_0 = 0.015$, 
and equal mean densities $\eta_{1,0}=\eta_{2,0}$ (excitation by 
vibrating walls).
From bottom to top: temperature profiles of both species,
density profiles $\eta_2(y)/(2\eta_0)$ and $\eta_1(y)/(2\eta_0)$.
Since $\sigma_1=\sigma_2$, the packing fraction $\eta_i$ is proportional to the
local density $\rho_i$ of species $i$.
The upper dashed curve shows the temperature ratio $\gamma = T_2/T_1$ as a 
function
of height, and the circles show the same quantity for a non equimolar mixture
where $\eta_{1,0} = 8 \eta_{2,0}$. \\
(b): Same with a higher mass ratio $m_1=5m_2$.
}
\label{fig:bina.85}
\end{figure}
\end{center}

\begin{center}
\begin{figure}[ht]
\epsfig{figure=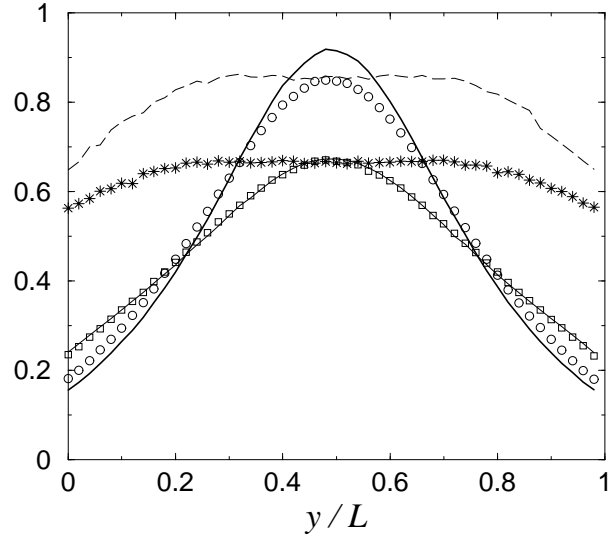,width=8cm,angle=0}
\caption{Density profiles and temperature ratio profiles (binary mixture, 
vibrating walls). The lines correspond to $\alpha_{ij} = 0.7; 0.8; 0.9$
whereas the symbols are associated with ``reverse'' inelasticities
$\alpha_{ij}=0.9;0.8;0.7$. The other parameters are 
$\alpha^t=0$, $m_1=3m_2$, $\eta_{1,0} = \eta_{2,0}$, 
$\eta_{0} = 2\eta_{1,0}=0.015$.
The upper flatter curves (dashed line and stars) 
show the temperature ratio. As in Fig.~\protect\ref{fig:bina.85}, 
the density of heavy particles $\rho_1$ 
(thick continuous curve and circles) is more
peaked and denser in the middle of the cell than that of light
grains (thin continuous curve and squares).
}
\label{fig:bin987789}
\end{figure}
\end{center}

\begin{center}
\begin{figure}[ht]
\epsfig{figure=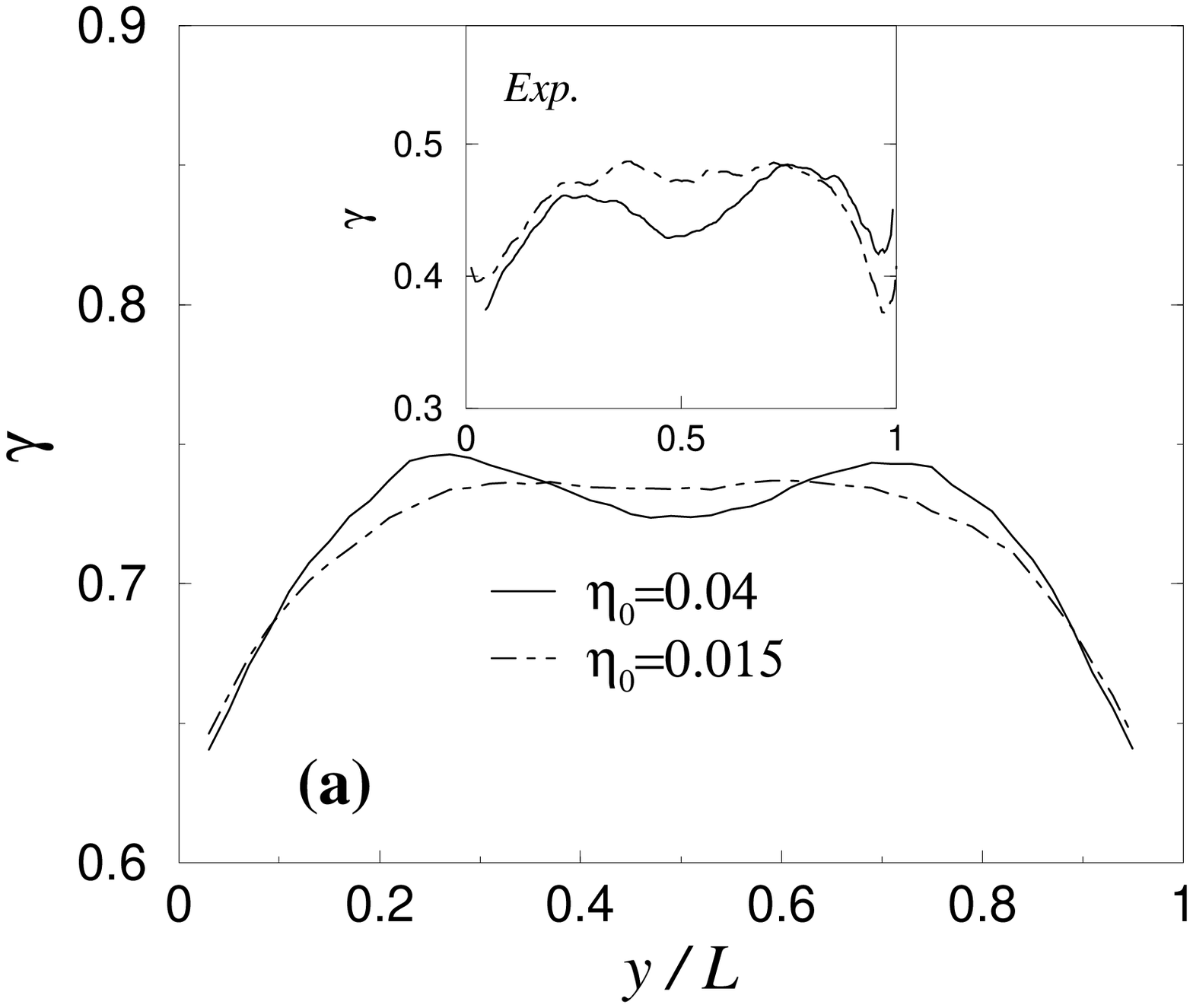,width=8cm,angle=0}
\epsfig{figure=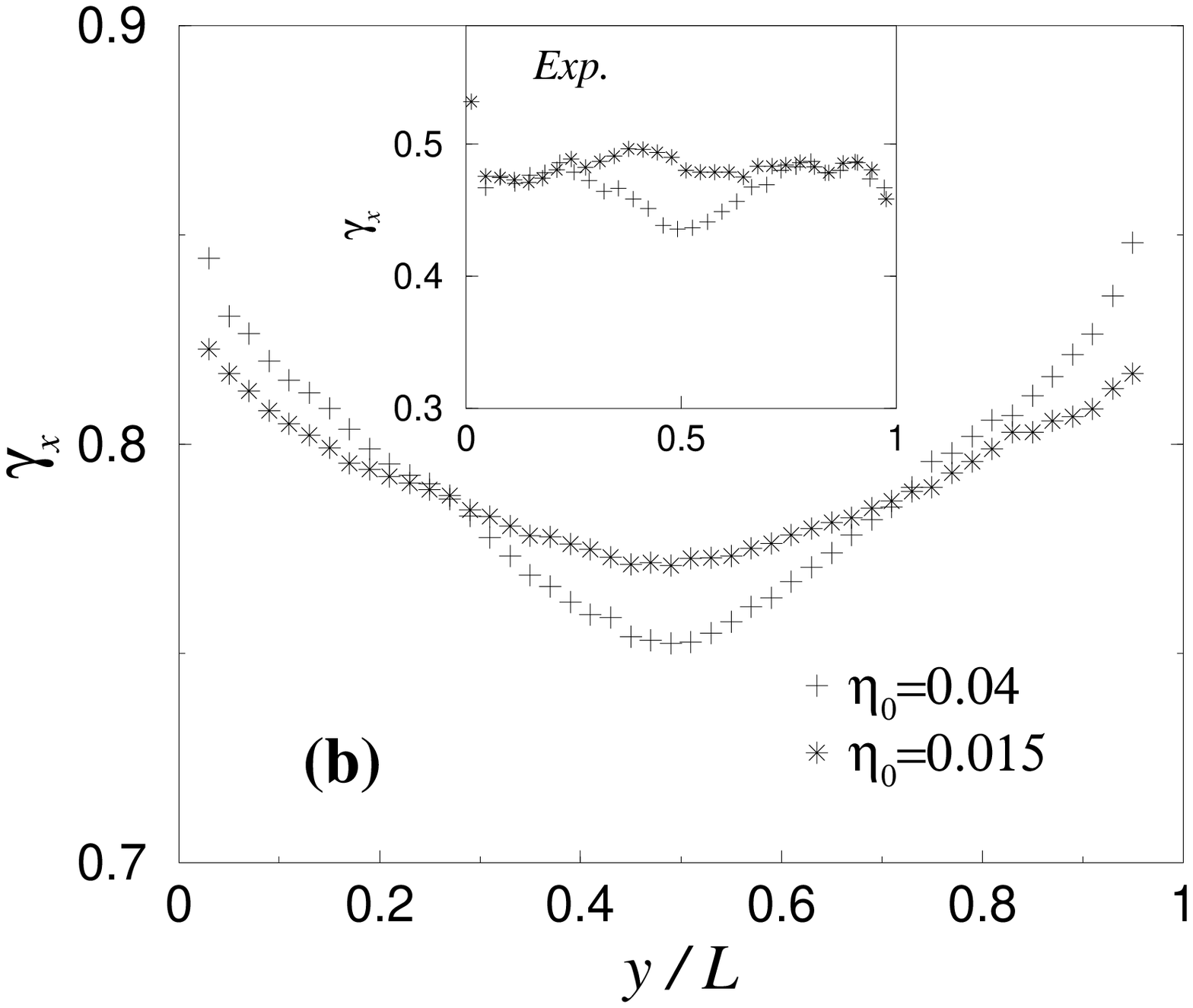,width=8cm,angle=0}
\caption{Effect of density on the temperature ratio for
$m_1=3m_2$, $\alpha_{ij}=0.9;0.8;0.7$ (vibrating walls).
Graph (a) shows the total ratio $T_2/T_1$ and 
graph (b) shows the ratio of horizontal temperatures
$T_{2,x}/T_{1,x}$. In both cases, the corresponding 
experimental measures are shown in the insets for a steel glass mixture
(at different densities, but with a density ratio of 2,
close to that of the MD simulations $0.04/0.015\simeq 2.6$). 
The purpose is to show that
the changes induced by density in MD are qualitatively
the same as in the experiments.
}
\label{fig:effetdens}
\end{figure}
\end{center}

\begin{center}
\begin{figure}[ht]
\epsfig{figure=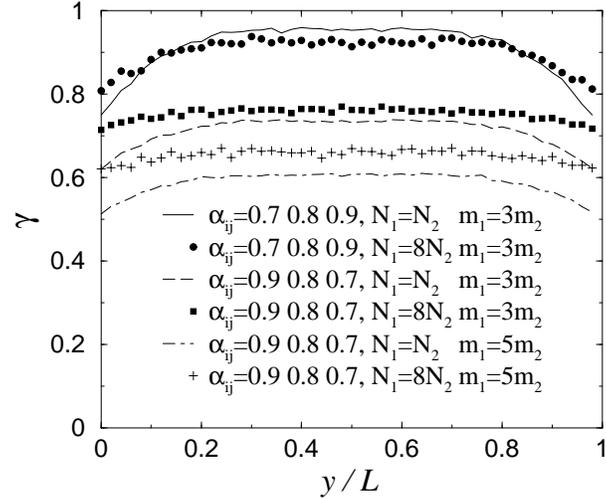,width=8cm,angle=0}
\caption{Influence of number fraction on the temperature
ratio $T_2/T_1$. The total number of particles is $N=N_1+N_2=500$ (vibrating
walls).
Given that $\sigma_1=\sigma_2$, $N_1/N_2=8$ corresponds to
$\eta_{1,0}= 8\eta_{2,0}$.}
\label{fig:effetfraction}
\end{figure}
\end{center}

The density and temperature profiles are displayed for various 
values of the parameters in 
Figs.~\ref{fig:bina.85} and \ref{fig:bin987789}.
The more massive particles (labeled 1), which display a more 
heterogeneous profile
and are denser in the middle of the cell, have typically larger kinetic
energies than the lighter ones: generically $\gamma=T_2/T_1$ is smaller than $1$,
as in homogeneous mixtures~\cite{garzo,Equipart}.
The study of the $y$-dependence of $\gamma$ shows that
$\gamma$ increases from the 
boundaries to the center of the system, and is constant across
a wide range of $y$ even if $T_1$ and $T_2$ vary significantly.
As also experimentally shown in~\cite{menon},
$\gamma$ is very close to $1$ if $m_1=m_2$, even
if the inelasticities of the particles are different. It 
decreases if the mass ratio increases (Figs.~\ref{fig:bina.85}), but
displays only a very weak (but strikingly similar to experimental data)
sensitivity on the global density (Fig.~\ref{fig:effetdens})
as well as on the relative densities of heavy and light particles;
moreover, $\gamma$ may increase or decrease as $\eta_{1,0}/\eta_{2,0}$
is increased (see Fig.~\ref{fig:effetfraction}), depending
on the relative inelasticities.

The anisotropy in the temperatures yield an anisotropic
$\gamma$; we obtain, as in experiments~\cite{Feitosa},
$\gamma_x > \gamma > \gamma_y$, with also different shapes: $\gamma_x$
decreases from the walls to the center while $\gamma$ and $\gamma_y$
increase (Fig.~\ref{fig:effetdens}).
All these results are in very good agreement with the existing experimental
results for two-dimensional vibrated mixtures~\cite{menon,Feitosa}.
We summarize in tables \ref{table:1} and \ref{table:2} some of the 
effects reported here.

When rotations are included (and thus $\alpha^t > -1$), $\gamma$ decreases.
Moreover, the ratio of rotational kinetic energies $\gamma_r$ can then
be measured. As shown in tables \ref{table:1} and \ref{table:2},
$\gamma_r$ takes values 
of the same order as $\gamma$. This quantity may also
be computed from experimental data, although measures of rotational velocities
are {\it a priori} more difficult than that of translational ones.

The measured values of $\gamma$ are of the same order as the experimental
data. We do not however try to obtain a precise numerical
agreement for the following
reasons: (i) in the experiments of \cite{menon}, the beads can rotate
in three dimensions, whereas the simulated spheres rotate in two dimensions
only; since $\alpha^t$ has a strong effect on $\gamma$, we suspect that
this difference between experiments and simulations may affect
$\gamma$; moreover, the experimental value of $\alpha^t$ is not known,
and the precise validity of the inelastic hard sphere model 
with a tangential restitution coefficient should be assessed;
(ii) different energy injection mechanisms (thermal vs. vibrating walls,
homogeneous driving vs. injection at the boundaries)
lead to different values of $\gamma$; even if the energy injection by
vibrating walls is reasonably realistic, such a sensitivity of
$\gamma$ renders its precise numerical prediction elusive.

Nonetheless, the {\em qualitative} very good agreement, 
even for subtle effects (see e.g. Fig.~\ref{fig:effetdens}),
between numerics and experiments, and the
possibility to change the various parameters in the simulations,
allow us to make some predictions on the effect of various
parameters: for example, increasing the mass ratio should yield
smaller values of $\gamma$ (Fig.~\ref{fig:bina.85}).
Moreover, Fig.~\ref{fig:bin987789} makes it clear that
the value of $\gamma$, at given mass ratio, is 
smaller for inelasticities $\alpha_{ij}=0.9; 0.8; 0.7$ than
with ``reverse'' inelasticities $\alpha_{ij}=0.7; 0.8; 0.9$. This effect
was already noted in~\cite{Equipart} and has the following intuitive
interpretation:
when the more massive particles are more inelastic, they loose more energy,
their temperature decreases which results in a higher $\gamma$.
We predict therefore that, in the context of the experiments reported 
in~\cite{menon},
a mixture of steel and aluminum ($\alpha_{steel} \approx 0.9$,
$\alpha_{al} \approx 0.83$, $m_{steel} \approx 3 m_{al}$) 
should yield a smaller value
of $\gamma$ than the brass-glass mixture ($\alpha_{brass} \approx 0.8$,
$\alpha_{glass} \approx 0.9$, $m_{brass} \approx 3 m_{glass}$) for which
the measured $\gamma$ is close to $0.6 - 0.7$. The dependence of $\gamma$
upon number fraction $x_i=\rho_i/\rho$
may on the other hand be counter-intuitive: at a given mean density
$\rho_0$, an increase of the relative fraction $x_1$ of heavy 
particles leads to an increase of $\gamma$
when the heavy particles are the more elastic 
(see Fig. \ref{fig:effetfraction}).
This effect was also clearly observed for the homogeneously
heated mixture \cite{Equipart}. On the other hand, an increase of 
$x_1$ 
leads a relatively weak decrease of $\gamma$ when the heavier particles are the
less elastic,
whereas the opposite
(albeit also quite weak) trend could be observed in \cite{Equipart}.

\begin{table}[hbt]
\centerline{
\begin{tabular}{lcccc}
\hline
$\alpha^t$ & $\gamma$ & $\gamma_x$ & $\gamma_r$  \\
\hline
 -1    & 0.88   & 0.92  &  -       \\
 -0.5  & 0.825  & 0.89  & 0.83     \\
 0     & 0.79   & 0.86  & 0.8      \\
\end{tabular}
\hspace{1cm}
\begin{tabular}{lcccc}
\hline
$\alpha^t$ & $\gamma$ & $\gamma_x$ & $\gamma_r$  \\
\hline
 -1    & 0.79   & 0.845  &  -       \\
 -0.5  & 0.7    & 0.78  & 0.69     \\
 0     & 0.65   & 0.74  & 0.66     \\
\end{tabular}
}
\caption{Values of $\gamma$, $\gamma_x$, $\gamma_r$ in the middle of the
system for $N=500$, $\alpha_{ij}=0.85$, $\eta_{1,0}=\eta_{2,0}$,
$m_1=3m_2$ (left) and $m_1=5m_2$ (right) }
\label{table:1}
\end{table}

\begin{table}[hbt]
\centerline{
\begin{tabular}{lcccc}
\hline
$\alpha^t$ & $\gamma$ & $\gamma_x$ & $\gamma_r$  \\
\hline
 -1    & 0.735  & 0.775  &  -       \\
 -0.5  & 0.69   & 0.735  & 0.735     \\
 0     & 0.665  & 0.72   & 0.72      \\
\end{tabular}
\hspace{1cm}
\begin{tabular}{lcccc}
\hline
$\alpha^t$ & $\gamma$ & $\gamma_x$ & $\gamma_r$  \\
\hline
 -1    & 0.95  & 1.    &  -       \\
 -0.5  & 0.89  & 0.99  & 0.84     \\
 0     & 0.85  & 0.96  & 0.81     \\
\end{tabular}
}
\caption{Values of $\gamma$, $\gamma_x$, $\gamma_r$ in the middle of the
system for $N=500$, $\alpha_{ij}=0.9, 0.8, 0.7$ (left)
and $\alpha_{ij}=0.7, 0.8, 0.9$ (right), $m_1=3m_2$, $\eta_{1,0}=\eta_{2,0}$.
}
\label{table:2}
\end{table}

\subsection{Velocity distributions}

\begin{center}
\begin{figure}[ht]
\epsfig{figure=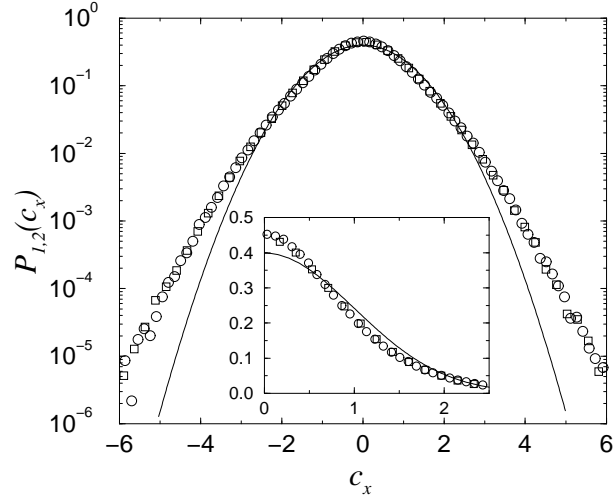,width=8cm,angle=0}
\caption{
Probability distribution functions  of the 
rescaled horizontal velocity components $c_{i,x} = v_{i,x}/\sqrt{T_{i,x}}$,
for an equimolar mixture. Squares are for $P_1$ (heavy grains) and circles 
for $P_2$ (light grains).
Here $\eta_0 = 0.015$, $N=500$, $\alpha_{ij}=0.9, 0.8, 0.7$,
$m_1=3m_2$ and $\alpha^t=0$.
The solid line is the Gaussian with variance $1$.}
\label{fig:pv1v2}
\end{figure}
\end{center}

As in the monodisperse case, we have measured the single-particle 
velocity distributions, which are anisotropic as expected. The vertical velocity
distributions are similar to those shown in Fig.~\ref{fig:pvy}, and
the horizontal velocity distributions show strong non-Gaussian
features, as in the monodisperse case. Moreover,
it appears in Fig.~\ref{fig:pv1v2} that the
rescaled velocity distributions $P_1(c_x)$ and $P_2(c_x)$ are
very close (even if not equal, see also~\cite{Pagnani}) 
for both types of particles.
The differences between $P_1(c_x)$ and $P_2(c_x)$ 
increase if the inelasticities or the mass ratio
increase. $P_i(c_x)$ depend slightly on the various parameters,
in the same way as the velocity distributions of the monodisperse
situation; 
this dependence would probably be very difficult to
measure in an experiment, which would probably
lead to the conclusion that $P_1(c_x) \approx P_2(c_x)$ .

\section{Conclusion}
\label{sec:concl}

In this study, we have considered vibrated granular gases well outside the
Boltzmann limit of (very) low densities. The molecular dynamics
simulations performed are free of the approximations underlying 
the usual kinetic theory or hydrodynamic approaches. Taking due account of
the first correction to the ideal gas contribution in the equation 
of state (second virial order), we however found a remarkable 
constant $yy$ component of the pressure tensor over the whole cell, for
monodisperse or bidisperse systems, despite
the strong density and temperature heterogeneities due to the realistic
energy injection mechanism. 

The study of the velocity distributions along the horizontal
direction (perpendicular to the energy injection) has revealed
non-Gaussian features similar to experiments, which depend weakly
on the various parameters involved in the model.

The projection of the dynamics onto the horizontal direction
has allowed us to gain insight into the correlations between
the effective restitution coefficient $\alpha_{1d}$ and
the relative velocities $g_x$ of colliding particles. The measured 
conditional probability distributions $\mu(\alpha_{1d} | g_x)$
are in agreement with the forms proposed in \cite{RRC2}, based upon
partial experimental data. The link between $\mu(\alpha_{1d} | g_x)$
and the velocity probability distribution functions 
\cite{RRC2} has been confirmed.

In the case of binary mixtures we have analyzed the ratio of
granular temperatures as a function of the various parameters, and found
a very good qualitative agreement with experiments. The velocity 
distributions of the two components have moreover been shown to be
very similar.

\acknowledgments
We would like to thank K. Feitosa and N. Menon for
providing us enlightening unpublished data. 

\appendix
\section{Inclusion of a tangential restitution coefficient}
\label{app:rot}

In this appendix we give the collision rules when a tangential
restitution coefficient is introduced (see also \cite{Luding}).
The two colliding particles, labeled $(1)$ and $(2)$, have masses
$m_i$, diameters $\sigma_i$, moment of inertia $I_i=m_i q \sigma_i/4$
with $q=1/2$ for disks and $2/5$ for spheres. The precolliding velocities
are ${\bm v}_i, {\bm \omega}_i$, and postcolliding velocities are denoted
with primes.

The normal unit vector is defined as:
\begin{equation}
\widehat{\bm \sigma} = \frac{ {\bm r}_1 -{\bm r}_2}{ |{\bm r}_1 -{\bm r}_2|} \ .
\end{equation}
The relative velocity of the contact point
\begin{equation}
{\bm g} = {\bm v}_1 - {\bm v}_2 - 
\left( \frac{\sigma_1}{2} {\bm \omega}_1 + \frac{\sigma_2}{2} {\bm \omega}_2
\right) \times \widehat{\bm \sigma} \ 
\end{equation}
has normal component 
${\bm g}_n = ({\bm g}\cdot \widehat{\bm \sigma}) \widehat{\bm \sigma}$ and
tangential component ${\bm g}_t = {\bm g} - {\bm g}_n$ (this defines
the tangential unit vector $\widehat{\bm t}= {\bm g}_t /|{\bm g}_t |$.

The postcollisional velocities can be expressed simply in terms of the
precollisional velocities through the introduction of the linear 
momentum change of particle $(1)$
\begin{equation}
\Delta {\bm P}=m_1 ({\bm v}_1' - {\bm v}_1)=-m_2({\bm v}_2' - {\bm v}_2) \ .
\end{equation}
Indeed the change of angular momentum is
\begin{equation}
\frac{2I_i}{\sigma_i} ({\bm \omega}_i'- {\bm \omega}_i)=
-\widehat{\bm \sigma} \times \Delta {\bm P}
\end{equation}
One obtains:
\begin{eqnarray}
{\bm v}_1'&=& {\bm v}_1 + \frac{\Delta {\bm P}}{m_1} \\
{\bm v}_2'&=& {\bm v}_2 - \frac{\Delta {\bm P}}{m_2} \\
{\bm \omega}_i' &=& {\bm \omega}_i -\frac{\sigma_i}{2I_i}
\widehat{\bm \sigma} \times \Delta {\bm P}
\end{eqnarray}
The normal and tangential components of $\Delta {\bm P}$ are then computed
using the definition of the normal and tangential coefficients of
restitution:
\begin{eqnarray}
{\bm g}_n'&=&-\alpha {\bm g}_n \\
{\bm g}_t'&=&-\alpha^t {\bm g}_t \ .
\end{eqnarray}
Since 
${\bm g}_n=[ ({\bm v}_1 -{\bm v}_2)\cdot 
\widehat{\bm \sigma}]\,\widehat{\bm \sigma}$, 
the first relation leads to
\begin{equation}
\Delta {\bm P} \cdot \widehat{\bm \sigma} = 
- \frac{m_1m_2}{m_1+m_2} (1+\alpha) ({\bm v}_1 -{\bm v}_2)\cdot 
\widehat{\bm \sigma} \ .
\end{equation}
Using the definition of ${\bm g}_t$, and with $I_i=m_i q \sigma_i/4$,
one obtains also
\begin{equation}
{\bm g}_t'={\bm g}_t + \Delta {\bm P}_t \left( \frac{1}{m_1}+\frac{1}{m_2}
\right)\left(1+\frac{1}{q}\right)
\end{equation}
[where 
$\Delta {\bm P}_t=(\Delta {\bm P} \cdot \widehat{\bm t})\widehat{\bm t}$].
Finally,
\begin{equation}
\Delta {\bm P} = - \frac{m_1m_2}{m_1+m_2} \left(
(1+\alpha) {\bm g}_n + \frac{1+\alpha^t}{1+1/q} {\bm g}_t \right)
\end{equation}

\section{Equation of state for a polydisperse inelastic mixture}
\label{app:eos}

In this appendix, we adopt a kinetic definition of the total pressure
and compute this quantity for an arbitrary homogeneous mixture of species $i$,
with number fraction $x_i = \rho_i/\rho$. 
Invoking the virial theorem,
the excess pressure 
$P^{\hbox{\scriptsize ex}}= P-P^{\hbox{\scriptsize ideal}} = 
P-\sum_i \rho_i T_i$ 
is related to the collisional transfer of
linear momentum: the partial excess pressure of species $i$ reads
(see e.g. \cite{Allen})
\begin{eqnarray}
P^{\hbox{\scriptsize ex}}_i &=& \lim_{t \to \infty} \, 
\frac{1}{d V} \,\frac{1}{t}\,
\sum_{j,\,\hbox{\scriptsize coll. partner of } i} 
\bm{r}_{ij}\cdot \delta {\bm p}_{i} \\
&=&  \lim_{t \to \infty} \, \frac{1}{d V} \,\frac{1}{t}\,
\sum_{j,\,\hbox{\scriptsize coll. partner of } i}
\frac{m_i\,m_j}{m_i+m_j} (1+\alpha_{ij}) 
(\widehat{\bm{\sigma}}\cdot \bm{v}_{ij})\,\sigma_{ij}
\quad \hbox{where}\quad \sigma_{ij} = \frac{\sigma_i+\sigma_j}{2}.
\label{eq:colltrans}
\end{eqnarray}
In these equations, it is understood that the summation runs 
over all the collision events
involving a particle of type $i$ and an arbitrary partner $j$, in a large
volume of measure $V$. The collisional transfer appearing in Eq. 
(\ref{eq:colltrans}) is readily computed within Enskog-Boltzmann 
kinetic theory, where the velocity distribution functions 
$\varphi_i(\bv)$ obey the set of non-linear
equations
\be
\partial_t \varphi_i ( \bm{v_1}, t) = \sum_{j=1}^{{\cal N}_s}\,
\chi_{ij} \sigma_{ij}^{d-1} \, n_j
\int d\bm{v}_2 \int d\widehat{\bm{\sigma}} \,
\Theta\left(\widehat{\bm{\sigma}}\cdot \bm{v}_{12}\right)
(\widehat{\bm{\sigma}}\cdot \bm{v}_{12})
\left[ \frac{1}{\alpha_{ij}^2}
\varphi_i(\bm{v}_1^\ast)\varphi_j(\bm{v}_2^\ast) - 
\varphi_i(\bm{v}_1)\varphi_j(\bm{v}_2) \right],
\ee 
where $\Theta$ denotes the Heavyside distribution and ($\bv^\ast_1,\bv^\ast_2$)
are the pre-collisional velocities converted into ($\bv_1,\bv_2$) by the
collision rule (\ref{eq:coll1})-(\ref{eq:coll2}). Equation (\ref{eq:colltrans})
may be rewritten
\be
P^{\hbox{\scriptsize ex}}_i \,=\, \frac{1}{2 d}\,\sum_{j=1}^{{\cal N}_s}
\chi_{ij} \sigma_{ij}^{d-1} \, n_j
\int d\bv_1d\bm{v}_2 \int d\widehat{\bm{\sigma}} \,
\Theta\left(\widehat{\bm{\sigma}}\cdot \bm{v}_{12}\right)
(\widehat{\bm{\sigma}}\cdot \bm{v}_{12})\,
\varphi_i(\bm{v}_1)\varphi_j(\bm{v}_2)\, 
\frac{m_i\,m_j}{m_i+m_j} (1+\alpha_{ij}) 
(\widehat{\bm{\sigma}}\cdot \bm{v}_{12})\,\sigma_{ij}.
\ee
Summing the contributions of all species, the total excess pressure follows:
\begin{eqnarray}
P^{\hbox{\scriptsize ex}} &=& \frac{1}{2 d}\,\sum_{i,j}
\chi_{ij}\, \sigma_{ij}^{d} \, n_i\,n_j\, \frac{m_i\,m_j}{m_i+m_j} 
(1+\alpha_{ij}) 
\int d\bv_1d\bm{v}_2 \int d\widehat{\bm{\sigma}} \,
\Theta\left(\widehat{\bm{\sigma}}\cdot \bm{v}_{12}\right)
(\widehat{\bm{\sigma}}\cdot \bm{v}_{12})^2\,
\varphi_i(\bm{v}_1)\varphi_j(\bm{v}_2) \\
&=& \frac{1}{2 d}\,\sum_{i,j}
\chi_{ij}\, \sigma_{ij}^{d} \, n_i\,n_j\, \frac{m_i\,m_j}{m_i+m_j} 
(1+\alpha_{ij})
\,\left[\int d\widehat{\bm{\sigma}} 
\Theta\left(\widehat{\bm{\sigma}}\cdot \widehat{\bm{v}}_{12}\right)
(\widehat{\bm{\sigma}}\cdot \widehat{\bm{v}}_{12})^2 \right]\,
\int d\bv_1d\bm{v}_2 \,(v_1^2+v_2^2)\,
\varphi_i(\bm{v}_1)\varphi_j(\bm{v}_2)
\label{eq:appdec}
\end{eqnarray}
where $\widehat{\bm{v}}_{12}$ is the unit vector along
$\bm{v}_{12}$, and
where the contribution from the dot product $\bv_1\cdot\bv_2$ vanishes
by symmetry in the last integral.  
The integral over the solid angle $\widehat{\bm\sigma}$ is related to the 
volume $V_d$ of a sphere with diameter 1:
\be
\int d\widehat{\bm{\sigma}} \,
\Theta\left(\widehat{\bm{\sigma}}\cdot \widehat{\bm{v}}_{12}\right)
(\widehat{\bm{\sigma}}\cdot \widehat{\bm{v}}_{12})^2 \,=\,
\frac{\pi^{d/2}}{d \,\Gamma(d/2)} \,=\, 2^{d-1} V_d,
\label{eq:B7}
\ee
where $\Gamma$ is the Euler function and it is understood that 
$\widehat{\bm{v}}_{12}$ denotes an arbitrary unit vector in (\ref{eq:B7}).
The volume 
$V_d$ is itself related to the packing fraction $\eta$ through
$\eta = \rho V_d \langle\sigma^d\rangle$. From the definition 
of kinetic temperatures $\int v^2 \varphi_i(\bv)\, d\bv = d T_i/m_i$,
we get 
\be
P^{\hbox{\scriptsize ex}}\,=\, \rho \,\eta \, 2^{d-2}\, 
\sum_{i,j}  \chi_{ij} \,
x_i x_j\, \frac{m_i m_j}{m_i+m_j} \, (1+\alpha_{ij}) \left(\frac{T_i}{m_i}
+ \frac{T_j}{m_j}\right) 
\frac{\sigma_{ij}^d}{\langle \sigma^d\rangle} 
\ee
from which we deduce the equation of state (\ref{eq:eos}). 
In this last step, no approximation (e.g. Gaussian etc) is made
concerning the $\varphi_i$. On the other
hand, the computation of any other moment 
$(\widehat{\bm{\sigma}}\cdot \bm{v}_{12})^p$ than $p = 2$ 
requires the detailed 
knowledge of the velocity distributions \cite{Pre2}. It is also 
noteworthy that the decoupling of velocities $\bv_1$ and $\bv_2$ in 
(\ref{eq:appdec}) is a specific property of the momentum transfer,
which significantly simplifies the calculation.

\section{Hydrodynamics}
\label{app:hydro}

In this appendix, we recall the hydrodynamical approach considered 
by Brey {\it et al.} \cite{Brey1}, and adapt it to the case of energy 
injection at both boundaries $y=0$ and $y=L$. The situation investigated in
\cite{Brey1} is that of a vibrating wall at $y=0$, and a reflecting wall
at $y=L$ so that the temperature and density gradients vanish at $y=L$. In our 
no-flow configuration with two vibrating walls, the gradients vanish by symmetry
in the middle of the cell ($y=L/2$), so that restricting to $y\in [0,L/2]$
allows to use directly the expressions derived in \cite{Brey1}
(which amounts to the formal identification $y\to 2y$ and $N \to N/2$. 
For completeness and clarity, we will however adapt the argument 
to our geometry.

In the case of a stationary system, without macroscopic velocity flow,
the hydrodynamic equations reduce to
\begin{eqnarray}
\nabla \cdot {\bm P} &=& 0 \\
\frac{2}{\rho d} \nabla \cdot {\bm q} + T \zeta &=& 0 \ .
\label{eq:c2}
\end{eqnarray}
Here ${\bm P}$ is the pressure tensor, ${\bm q}$ is the heat flux, and $\zeta$
the cooling rate due to the collisional energy dissipation.
In the Navier-Stokes approximation for a low density gas described by the
Boltzmann equation modified to account for the inelastic nature of collisions
\cite{IHSDynamics}, 
\begin{eqnarray}
{\bm P} &=& P {\bm I} \\
{\bm q} &=& -\kappa \nabla T - \mu \nabla \rho
\end{eqnarray}
where $P$ is the ideal gas pressure: $P= \rho T$.
The explicit expressions of the heat conductivity $\kappa$, the transport
coefficient $\mu$ and cooling rate $\zeta$ may 
be found in \cite{Brey1}. The important ingredient is that
$\mu$ is proportional to $T^{3/2}/\rho$ and $\kappa$ to
$\sqrt{T}$, while $\zeta \propto p/\sqrt{T}$, with coefficients
depending on the inelasticity $\alpha$.

The system is considered homogeneous in the $x$ direction, so that only
gradients along the $y$ direction are taken into account. We emphasize that
the ideal gas equation of state $P=\rho T$ is assumed, and this simplification
is an important ingredient in the following derivation. The previous 
equations then reduce to:
\begin{eqnarray}
\frac{\partial P}{\partial y} &=& 0 \\
\frac{2A(\alpha)}{d \rho} \frac{\partial}{\partial y}
\left( \sqrt{T} \frac{\partial T}{\partial y} \right)
- p \sqrt{T} &=&0.
\label{eq:t}
\end{eqnarray}
In order to simplify the equation on the temperature,
it is convenient to introduce a new variable $\xi$ defined by
\begin{equation}
d\xi =  \sqrt{a(\alpha)} \frac{dy}{\lambda(y)} = 
C \sigma^{d-1} \sqrt{a(\alpha)} \rho (y) dy
\end{equation}
where $\lambda (y)  = [C \sigma^{d-1} \rho (y)]^{-1}$ is the
mean-free-path ($C=2\sqrt{2}$ for $d=2$), and $a(\alpha)$
includes all the dependence in $\alpha$. Equation (\ref{eq:t})
now reads
\begin{equation}
\frac{\partial^2}{\partial \xi^2} \sqrt{T} = \sqrt{T} \ .
\end{equation}
The variable $\xi$ takes values between $0$ and $\xi_m$, 
with $\xi_m \propto N$.
Then $\sqrt{T} = A \exp(-\xi) + B \exp(\xi)$
where $A$ and $B$ depend on the boundary conditions.
In the case of two vibrating walls, the solution is symmetric with respect
to $y=L/2$ (or $\xi=\xi_m/2$). With $T(0)=T(\xi_m)=T_0$ one obtains
\begin{equation}
T(\xi) = \frac{T_0}{\sinh^2 \xi_m}
\left(
\sinh (\xi_m -\xi) + \sinh \xi
\right)^2.
\end{equation}
It is possible to integrate $d\xi = C \sigma^{d-1} \sqrt{a(\alpha)} n(y) dy=
C \sigma^{d-1}\sqrt{a(\alpha)} p dy/T(y)$ to obtain $y(\xi)$ and $P$:
\begin{eqnarray}
P &=& \frac{T_0}{2C \sigma^{d-1} L \sqrt{a(\alpha)} 
\cosh^2 \frac{\xi_m}{2} } 
\left( \xi_m + \sinh \xi_m \right) \\
\frac{y}{L}&=& \frac{\xi + \sinh \xi \cosh (\xi_m - \xi)}
{\xi_m + \sinh \xi_m}.
\end{eqnarray}
Those equations are the same as for the case of one vibrating wall
\cite{Brey1},
but with $\xi_m \rightarrow  2\xi_m$ and $L  \rightarrow  2L$,
as expected on the basis on the symmetry argument proposed above.
It is possible to invert $T(\xi)$ and therefore to obtain the 
profiles $y(T)$ (two symmetric branches):
\begin{eqnarray}
\xi &=& \frac{\xi_m}{2} \pm \cosh^{-1} \left( \sqrt{\frac{T}{T_0}} 
\cosh \frac{\xi_m}{2} \right) \\
\frac{y}{L}&=& \frac{\xi + \sinh \xi \cosh (\xi_m - \xi)}
{\xi_m + \sinh \xi_m}  \ .
\end{eqnarray}



\newpage

\end{document}